\documentclass{ieeeaccess}
\usepackage{cite}
\usepackage{amsmath,amssymb,amsfonts}
\usepackage{graphicx}
\usepackage{textcomp}
\usepackage{pdfrender}
\usepackage{url}

\usepackage{soul}
\sethlcolor{yellow}

\usepackage{amssymb}

\usepackage[noend]{algorithmic}
\usepackage{algorithm}

\usepackage{multirow}

\usepackage{bm}
\makeatletter
\AtBeginDocument{\DeclareMathVersion{bold}
\SetSymbolFont{operators}{bold}{T1}{times}{b}{n}
\SetSymbolFont{NewLetters}{bold}{T1}{times}{b}{it}
\SetMathAlphabet{\mathrm}{bold}{T1}{times}{b}{n}
\SetMathAlphabet{\mathit}{bold}{T1}{times}{b}{it}
\SetMathAlphabet{\mathbf}{bold}{T1}{times}{b}{n}
\SetMathAlphabet{\mathtt}{bold}{OT1}{pcr}{b}{n}
\SetSymbolFont{symbols}{bold}{OMS}{cmsy}{b}{n}
\renewcommand\boldmath{\@nomath\boldmath\mathversion{bold}}}

\DeclareFontShape{T1}{formata}{m}{sl}{<->ssub*formata/m/it}{}
\makeatother

\def\BibTeX{{\rm B\kern-.05em{\sc i\kern-.025em b}\kern-.08em
    T\kern-.1667em\lower.7ex\hbox{E}\kern-.125emX}}

\begin{document}
\history{Date of publication XXX.}
\doi{XXX}

\title{Joint Hardware-Workload Co-Optimization for In-Memory Computing Accelerators}
\author{\uppercase{Olga Krestinskaya}\authorrefmark{1}, 
\uppercase{Mohammed E. Fouda}\authorrefmark{2}, \uppercase{Ahmed Eltawil}
\authorrefmark{1}, and \uppercase{Khaled N. Salama}\authorrefmark{1}
}

\address[1]{King Abdullah University of Science and Technology (KAUST), Thuwal, Saudi Arabia (e-mails: ok@ieee.org,  ahmed.eltawil@kaust.edu.sa, khaled.salama@kaust.edu.sa)}
\address[2]{Compumacy for Artificial Intelligence Solutions, Cairo, Egypt (e-mail: foudam@uci.edu)}
\tfootnote{This work was supported by the King Abdullah University of Science and Technology through the Competitive Research Grant program under grant  URF/1/4704-01-01.}

\markboth
{Author \headeretal: Joint Hardware-Workload Co-Optimization for In-Memory Computing Accelerators}
{Author \headeretal: Joint Hardware-Workload Co-Optimization for In-Memory Computing Accelerators}


\begin{abstract}

Software-hardware co-design is essential for optimizing in-memory computing (IMC) hardware accelerators for neural networks. However, most existing optimization frameworks target a single workload, leading to highly specialized hardware designs that do not generalize well across models and applications. In contrast, practical deployment scenarios require a single IMC platform that can efficiently support multiple neural network workloads.
This work presents a joint hardware-workload co-optimization framework based on an optimized evolutionary algorithm for designing generalized IMC accelerator architectures. By explicitly capturing cross-workload trade-offs rather than optimizing for a single model, the proposed approach significantly reduces the performance gap between workload-specific and generalized IMC designs.
The framework is evaluated on both RRAM- and SRAM-based IMC architectures, demonstrating strong robustness and adaptability across diverse design scenarios. Compared to baseline methods, the optimized designs achieve energy-delay-area product (EDAP) reductions of up to 76.2\% and 95.5\% when optimizing across a small set (4 workloads) and a large set (9 workloads), respectively. The source code of the framework is available at https://github.com/OlgaKrestinskaya/JointHardwareWorkloadOptimizationIMC.

\end{abstract}


\titlepgskip=-21pt

\maketitle

\section{Introduction}
\label{sec:introduction}
\PARstart{W}{ith}  {the rapid growth in neural network models and widespread AI adoption, optimized and energy-efficient hardware has become essential to sustain AI progress. Traditional von Neumann architectures are often inefficient for AI workloads, motivating the development of specialized AI accelerators tailored for neural network execution. In-memory computing (IMC) is a promising approach that integrates memory and computation, reducing data movement between processing units and off-chip memory} \cite{wan2022compute, sebastian2020memory, fouda2022memory, krestinskaya2024neural, yantir2022hardware, jain2019neural, le202364, mannocci2023memory}.  {This leads to significant gains in energy efficiency, performance, and throughput} \cite{zhang2020neuro, smagulova2023resistive, aguirre2024hardware, krestinskaya2022towards}.  {Beyond high-performance accelerators, IMC is well suited for low-power, on-edge applications, enabling intelligent systems outside the cloud} \cite{dalgaty2024mosaic}.

Since in-memory computing (IMC) accelerators are often fine-tuned specifically for neural network applications,  {software-hardware co-design is essential for achieving optimal performance} \cite{yu2025full, kim2025efficient}.  {This requires simultaneous optimization of neural network models, mapping strategies, and hardware parameters across device, circuit, architecture, and system levels} \cite{krestinskaya2024neural,  krestinskaya2023towards}. 
 {Hardware design space exploration and hardware-aware neural architecture search are promising co-design approaches that automate this process and identify optimal network and hardware configurations using dedicated search algorithms} \cite{krestinskaya2024neural, krestinskaya2025cimnas}.

\begin{figure}[t!]
    \includegraphics[width=\columnwidth]{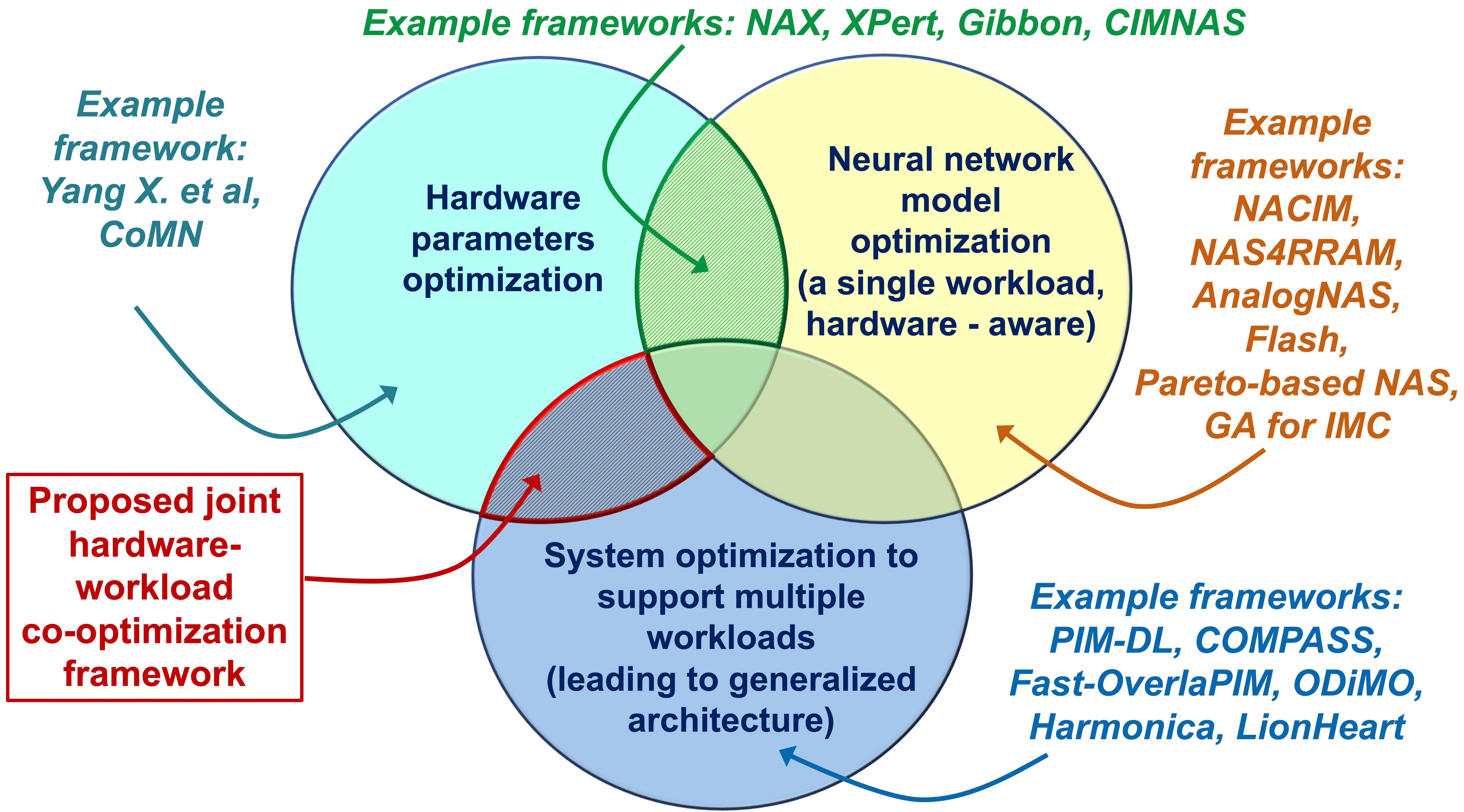}
    \caption{Research gap and representative state-of-the-art frameworks (including \cite{negi2022nax, sun2023gibbon, moitra2023xpert, yang2021multi, yuan2021nas4rram, han2024comn, benmeziane2023analognas, krestinskaya2020towards, li2021flash, guan2022hardware, jiang2020device, krestinskaya2020automating, zhou2021pim, park2025compass, wang2024fast, risso2023precision, behnam2024harmonica, lammie2025lionheart, krestinskaya2025cimnas}, which are discussed in detail in Section~\ref{Sback}).}
    \label{fmotive}
\end{figure}

 {Most of the state-of-the-art frameworks for such co-design focus on a single neural network model (i.e., a single workload), either by incorporating hardware feedback during model design or by optimizing hardware parameters alone} \cite{guan2022hardware, krestinskaya2020automating, yuan2021nas4rram, jiang2020device, benmeziane2023analognas, krestinskaya2020towards, li2021flash, yang2021multi}.  {Some approaches co-optimize network and hardware configurations} \cite{negi2022nax, sun2023gibbon, moitra2023xpert},  {but they remain model-specific and restricted to a single workload.
Other studies optimize the mapping of diverse workloads onto fixed IMC hardware} \cite{zhou2021pim, park2025compass, wang2024fast, risso2023precision, behnam2024harmonica, lammie2025lionheart}.  {However, optimizing IMC hardware parameters to efficiently support multiple workloads remains an open challenge, which this study addresses, as illustrated in} Fig.\ref{fmotive}.

 {In this study, we introduce a hardware-workload co-optimization framework for IMC-based neural network accelerators that targets a generalized IMC architecture capable of efficiently supporting diverse workloads. To mitigate the efficiency loss typically associated with hardware generality, the proposed method aims to minimize the performance gap between generalized IMC designs and hardware optimized for a single workload.
The framework relies on a four-phase genetic algorithm (GA) with an optimized Hamming-distance-based sampling strategy to address convergence limitations of state-of-the-art methods. In addition, it explores a broad and large-scale hardware design space spanning device, circuit, architecture, and system levels, which is often overlooked in existing approaches.
We validate the framework on RRAM- and SRAM-based IMC architectures under multiple objectives, demonstrating adaptability across diverse system constraints. We further perform hardware-workload-technology co-optimization to reveal performance cost trade-offs across CMOS nodes and confirm scalability on a broader set of workloads.}

\begin{table*}[t!]
\centering
\caption{Comparison of the proposed framework with state-of-the-art approaches.}
\resizebox{2.05\columnwidth}{!}{%
\begin{tabular}{|lccccccccc|}
\hline
\multicolumn{1}{|c|}{\multirow{3}{*}{\begin{tabular}[c]{@{}c@{}}\textbf{Frame-}\\ \textbf{work}\end{tabular}}}         & \multicolumn{1}{c|}{\multirow{3}{*}{\textbf{Algorithm$^{*1}$}}}          & \multicolumn{1}{c|}{\multirow{3}{*}{\begin{tabular}[c]{@{}c@{}}\textbf{Models} \\  \textbf{tested}\end{tabular}}} & \multicolumn{1}{c|}{\multirow{3}{*}{\begin{tabular}[c]{@{}c@{}}\textbf{Hardware} \\ \textbf{search space} \\ \textbf{size} $^{*2}$\end{tabular}}} & \multicolumn{4}{c|}{\textbf{Optimization parameters}}                                                                                                                                                                                                                                                                                                                                                                                                                  & \multicolumn{1}{c|}{\multirow{3}{*}{\begin{tabular}[c]{@{}c@{}}\textbf{Work-}\\ \textbf{load}\end{tabular}}} & \multirow{3}{*}{\begin{tabular}[c]{@{}c@{}}\textbf{Generalized} \\ \textbf{hardware} \\ \textbf{solution$^{*7}$}\end{tabular}} \\ \cline{5-8}
\multicolumn{1}{|c|}{}                                                                                        & \multicolumn{1}{c|}{}                                                    & \multicolumn{1}{c|}{}                                                                                            & \multicolumn{1}{c|}{}                                                                                                           & \multicolumn{1}{c|}{\multirow{2}{*}{\textbf{D$^{*3}$}}}                       & \multicolumn{1}{c|}{\multirow{2}{*}{\textbf{C$^{*4}$}}}                                                                             & \multicolumn{1}{c|}{\multirow{2}{*}{\textbf{A$^{*5}$}}}                                                                                    & \multicolumn{1}{c|}{\multirow{2}{*}{\textbf{S$^{*6}$}}}                                             & \multicolumn{1}{c|}{}                                                                               &                                                                                                              \\
\multicolumn{1}{|c|}{}                                                                                        & \multicolumn{1}{c|}{}                                                    & \multicolumn{1}{c|}{}                                                                                            & \multicolumn{1}{c|}{}                                                                                                           & \multicolumn{1}{c|}{}                                                         & \multicolumn{1}{c|}{}                                                                                                               & \multicolumn{1}{c|}{}                                                                                                                      & \multicolumn{1}{c|}{}                                                                               & \multicolumn{1}{c|}{}                                                                               &                                                                                                              \\ \hline
\multicolumn{1}{|l|}{{\begin{tabular}[c]{@{}l@{}}NAX$^{*8}$ \\ \cite{negi2022nax}\end{tabular}}}       & \multicolumn{1}{c|}{DS}                                                  & \multicolumn{1}{c|}{ResNet}                                                                                      & \multicolumn{1}{c|}{\begin{tabular}[c]{@{}c@{}}$4.2\times10^6$\\ (layer-wise)\end{tabular}}                                     & \multicolumn{1}{c|}{-}                                                        & \multicolumn{1}{c|}{\begin{tabular}[c]{@{}c@{}}Crossbar size\end{tabular}}                                                       & \multicolumn{1}{c|}{-}                                                                                                                     & \multicolumn{1}{c|}{-}                                                                              & \multicolumn{1}{c|}{single}                                                                         & -                                                                                                            \\ \hline
\multicolumn{1}{|l|}{{\begin{tabular}[c]{@{}l@{}}Yang X. \\ et al. \cite{yang2021multi}\end{tabular}}} & \multicolumn{1}{c|}{\begin{tabular}[c]{@{}c@{}}CF-\\ MESMO\end{tabular}} & \multicolumn{1}{c|}{\begin{tabular}[c]{@{}c@{}}ResNet, \\ VGG\end{tabular}}                                      & \multicolumn{1}{c|}{$1.35\times10^2$}                                                                                           & \multicolumn{1}{c|}{\begin{tabular}[c]{@{}c@{}}Bits \\ per cell\end{tabular}} & \multicolumn{1}{c|}{Crossbar size}                                                                                                  & \multicolumn{1}{c|}{-}                                                                                                                     & \multicolumn{1}{c|}{\begin{tabular}[c]{@{}c@{}}Frequency,\\ temperature\end{tabular}}               & \multicolumn{1}{c|}{single}                                                                         & -                                                                                                            \\ \hline
\multicolumn{1}{|l|}{{\begin{tabular}[c]{@{}l@{}}Gibbon \\ \cite{sun2023gibbon}\end{tabular}}}         & \multicolumn{1}{c|}{EA}                                                  & \multicolumn{1}{c|}{\begin{tabular}[c]{@{}c@{}}ResNet \\ (modified)\end{tabular}}                                & \multicolumn{1}{c|}{$6.4\times10^1$}                                                                                            & \multicolumn{1}{c|}{\begin{tabular}[c]{@{}c@{}}Bits \\ per cell\end{tabular}} & \multicolumn{1}{c|}{\begin{tabular}[c]{@{}c@{}}Crossbar size,\\ ADC/DAC resolution\end{tabular}}                                    & \multicolumn{1}{c|}{-}                                                                                                                     & \multicolumn{1}{c|}{-}                                                                              & \multicolumn{1}{c|}{single}                                                                         & -                                                                                                            \\ \hline
\multicolumn{1}{|l|}{{\begin{tabular}[c]{@{}l@{}}XPert$^{*8}$ \\ \cite{moitra2023xpert}\end{tabular}}} & \multicolumn{1}{c|}{DS}                                                  & \multicolumn{1}{c|}{VGG}                                                                                         & \multicolumn{1}{c|}{\begin{tabular}[c]{@{}c@{}}$4.5\times10^{15}$ \\ (layer-wise)\end{tabular}}                                 & \multicolumn{1}{c|}{-}                                                        & \multicolumn{1}{c|}{\begin{tabular}[c]{@{}c@{}}Column sharing,\\ ADC precision/type\end{tabular}}                                   & \multicolumn{1}{c|}{-}                                                                                                                     & \multicolumn{1}{c|}{-}                                                                              & \multicolumn{1}{c|}{single}                                                                         & -                                                                                                            \\ \hline
\multicolumn{1}{|l|}{{\begin{tabular}[c]{@{}l@{}}CoMN \\ \cite{han2024comn}\end{tabular}}}             & \multicolumn{1}{c|}{BO}                                                  & \multicolumn{1}{c|}{\begin{tabular}[c]{@{}c@{}}ResNet, \\ VGG, \\ MobileNet\end{tabular}}                        & \multicolumn{1}{c|}{$2.2\times10^4$}                                                                                            & \multicolumn{1}{c|}{-}                                                        & \multicolumn{1}{c|}{\begin{tabular}[c]{@{}c@{}}Crossbar size, \\ number of ADCs, \\ ADC resolution\\ /power/frequency\end{tabular}} & \multicolumn{1}{c|}{\begin{tabular}[c]{@{}c@{}}Crossbars per tile, \\ buffer size\\ /bandwidth, \\ router flit bandwidth\end{tabular}}     & \multicolumn{1}{c|}{Mapping}                                                                        & \multicolumn{1}{c|}{single}                                                                         & -                                                                                                            \\ \hline
\multicolumn{1}{|l|}{{\begin{tabular}[c]{@{}l@{}}CIMNAS \\ \cite{krestinskaya2025cimnas}\end{tabular}}}             & \multicolumn{1}{c|}{EA}                                                  & \multicolumn{1}{c|}{\begin{tabular}[c]{@{}c@{}}ResNet, \\ MobileNet\end{tabular}}                        & \multicolumn{1}{c|}{$1.4\times10^7$}                                                                                            & \multicolumn{1}{c|}{\begin{tabular}[c]{@{}c@{}}Bits \\ per cell\end{tabular}} & \multicolumn{1}{c|}{Crossbar size}                                                                                                  & \multicolumn{1}{c|}{\begin{tabular}[c]{@{}c@{}}Crossbars per tile, \\ buffer size, \\tiles per router,\\ tile groups\\ \end{tabular}}  & \multicolumn{1}{c|}{\begin{tabular}[c]{@{}c@{}}Frequency,\\ operating \\voltage\end{tabular}}                                                                        & \multicolumn{1}{c|}{single}                                                                         & -                                                                                                            \\ \hline
\multicolumn{1}{|l|}{\textbf{Ours}}                                                                           & \multicolumn{1}{c|}{\begin{tabular}[c]{@{}c@{}}Improved \\EA \end{tabular}}                                                   & \multicolumn{1}{c|}{\begin{tabular}[c]{@{}c@{}}ResNet, AlexNet, \\VGG, MobileNet, \\ DenseNet, ViT, \\ MobileBERT,  \\GPT-2 Medium$^{*9}$\end{tabular}}            & \multicolumn{1}{c|}{$1.21\times10^7$}                                                                                           & \multicolumn{1}{c|}{\begin{tabular}[c]{@{}c@{}}Bits \\ per cell\end{tabular}} & \multicolumn{1}{c|}{Crossbar size}                                                                                                  & \multicolumn{1}{c|}{\begin{tabular}[c]{@{}c@{}}Crossbars per tile, \\ tiles per router,\\ tile groups per chip\\ buffer size\end{tabular}} & \multicolumn{1}{c|}{\begin{tabular}[c]{@{}c@{}}Cycle time$^{*10}$,\\ operating \\voltage, \\ technology$^{*11}$\end{tabular}} & \multicolumn{1}{c|}{\textbf{multiple}}                                                              & \checkmark                                                                                                   \\ \hline
\multicolumn{10}{|l|}{\begin{tabular}[c]{@{}l@{}}$^{*1}$: DS - differential search, CF-MESMO - continuous-fidelity max-value entropy search for multi-objective optimization, EA - evolutionary algorithm,  BO - Bayesian \\ optimization. $^{*2}$: calculated based on the hardware search space only. $^{*3}$: Device optimization, $^{*4}$: Circuit optimization,  $^{*5}$: Architecture optimization, \\ $^{*6}$: Additional system-related parameters. $^{*7}$: Generalized hardware solution output supporting different workloads. $^{*8}$: Large hardware search space due to different \\ crossbar  sizes in each layer. $^{*9}$: Varies based on the experiment. $^{*10}$: Equivalent to frequency, $^{*11}$: used in some experiments\end{tabular}}                                                                                                                                                                                                                                                                                                                                                                                                                                       \\ \hline
\end{tabular}
}
\label{t1}
\end{table*}

 {The main contributions of this work are:}
\begin{itemize}
\item  {A hardware-workload co-optimization framework for generalized IMC architectures that efficiently supports multiple neural network workloads (in contrast to state-of-the-art single-workload optimization approaches).}
\item  {A four-phase genetic algorithm with Hamming-distance-based sampling that improves convergence and yields consistently high-quality hardware designs.}
\item  {A comprehensive optimization across device, circuit, architecture, and system levels (unlike state-of-the-art approaches with limited search spaces).}
\item  {A demonstration that the performance gap between generalized and workload-specific IMC designs can be minimized without sacrificing efficiency.}
\end{itemize}

The paper is organized as follows: Section~\ref{Sback} reviews background and related work. Section~\ref{Sproposed} details the proposed framework, algorithm, and hardware search space. Section~\ref{results} presents simulation results, highlighting convergence improvements, performance gains, trade-offs between RRAM and SRAM configurations, as well as practical scenarios and scalability. Section~\ref{Sdisc} discusses the advantages, practical applications,  {limitations}, and  {possibilities to extend the proposed method}. Section~\ref{Sconclu} concludes the paper.

\section{Background}
\label{Sback}

\subsection{ {Software-hardware co-design for AI applications}}

 {Software-hardware co-design is necessary for efficient AI system development, as modern workloads increasingly require tight coupling between neural network models and the underlying computing platforms. AI hardware, including GPUs, FPGAs, domain-specific accelerators, and IMC solutions, achieves high performance and energy efficiency by jointly optimizing model architectures, numerical precision, dataflow, and hardware resources, rather than treating algorithms and hardware as independent design layers} \cite{waheed2025edge, sabih2024hardware, jameil2022efficient, an2024fire, guo2025survey}.  {For IMC specifically, achieving high efficiency requires adapting neural network models to the constraints and non-idealities of IMC hardware, while simultaneously optimizing the hardware to efficiently support target workloads} \cite{yu2025full, bai2024end}.  {Often, such optimization involves a large number of design parameters, including neural network configurations and hardware architectural choices. The strong interdependence between model characteristics and IMC design choices further leads to extremely large and complex search spaces, making manual exploration and tuning impractical} \cite{krestinskaya2024neural, krestinskaya2025cimnas}.  {Consequently, automated hardware design space exploration and hardware-aware neural architecture search methods become essential for efficient and systematic IMC co-optimization} \cite{zhang2024easyacim, lv2025situ, krestinskaya2024neural}.

\subsection{ {Automated co-design frameworks for IMC}}

The majority of state-of-the-art methods for optimizing IMC hardware for neural networks fall into three categories: (1) neural network model optimization with IMC hardware feedback (typically for a single workload), (2) IMC hardware parameter optimization for a specific workload, and (3) co-optimization of both model and hardware parameters \cite{krestinskaya2024neural, krestinskaya2024towards}.
Frameworks focusing on neural network model optimization with fixed hardware, such as NACIM \cite{jiang2020device}, NAS4RRAM \cite{yuan2021nas4rram}, AnalogNAS \cite{benmeziane2023analognas}, Flash \cite{li2021flash}, Pareto-based NAS \cite{guan2022hardware}, and GA-based approaches \cite{krestinskaya2020automating, krestinskaya2020towards}, fine-tune model parameters using feedback from IMC hardware performance.  {These approaches optimize AI model configurations, such as the number of blocks, their interconnections, and kernel sizes, while incorporating hardware metrics into the optimization loop to maximize efficiency} \cite{guan2022hardware},  {or take hardware feedback into account to mitigate and adapt to hardware non-idealities} \cite{krestinskaya2020automating, krestinskaya2020towards}.

In contrast, IMC hardware parameter optimization techniques, often referred to as design space exploration methods, target the selection of optimal hardware parameters for a given model. For example, the approach in \cite{yang2021multi} focuses on hardware optimization for a fixed neural network  {using a differential search (DS) method for a relatively small search space.} 
 {The other framework, CoMN} \cite{han2024comn},  {leverages Bayesian Optimization (BO) to tune circuit- and architecture-level parameters, yet the hardware search space remains relatively small. }

Model-hardware co-optimization methods simultaneously optimize both neural network and hardware parameters, typically fine-tuning hardware for a single optimized model, as seen in Gibbon \cite{sun2023gibbon}, CIMNAS \cite{krestinskaya2025cimnas}, NAX \cite{negi2022nax}, and XPert \cite{moitra2023xpert}.
NAX \cite{negi2022nax}  {targets kernel size and crossbar size optimization using a Differential Search (DS) method to enhance energy efficiency.  Gibbon} \cite{sun2023gibbon}  {co-optimizes neural network parameters and hardware using an Evolutionary Algorithm (EA), and it primarily focuses on model structure and quantization optimization, and the hardware search space is relatively limited. CIMNAS \cite{krestinskaya2025cimnas} also uses EA to explore an expanded design space comprising up to $10^{85}$ parameter combinations through the joint optimization of model, quantization and IMC hardware parameters.
XPert} \cite{moitra2023xpert}  {uses a two-stage optimization strategy. It first explores hardware-level parameters such as channel depth, ADC type, and column sharing to reduce latency and on-chip area. Then, it fine-tunes input and ADC precision for energy and accuracy trade-offs. This framework supports a relatively large search space through a layer-wise optimization scheme, where different hardware configurations are applied for different layers.
While focusing on hardware parameter optimization, the aforementioned frameworks target a single workload or different workloads separately.}

\subsection{ {Multiple workloads support}}

The optimization of systems to support multiple workloads can be viewed as efficiently mapping diverse neural network workloads to fixed IMC accelerators (Fig.~\ref{fmotive}). Existing approaches generally fall into two categories: (i) mapping-centric methods that optimize IMC accelerators for improved data transfer and computational speed, and (ii) heterogeneous systems that combine analog and digital cores with conventional processors to handle diverse workloads. Frameworks such as PIM-DL \cite{zhou2021pim}, COMPASS \cite{park2025compass}, and Fast-OverlaPIM \cite{wang2024fast} belong to the first category, where PIM-DL optimizes data layout for digital IMC accelerators by considering parallelism, memory use, and movement costs; COMPASS uses a genetic algorithm-based compiler to partition large networks across fixed IMC hardware while reducing IMC-DRAM transfers; and Fast-OverlaPIM accelerates consecutive layer processing by exploiting data overlap. The second category includes frameworks such as ODiMO \cite{risso2023precision}, Harmonica \cite{behnam2024harmonica}, and LionHeart \cite{lammie2025lionheart}, where ODiMO partitions networks across analog and digital accelerators using gradient-based optimization, Harmonica applies Hessian-sensitivity analysis to split layers between IMC and digital cores while preserving accuracy, and LionHeart co-optimizes mappings to account for non-idealities in analog IMC units. While these frameworks focus on mapping workloads to fixed hardware, the joint workload-hardware co-optimization proposed in this work instead optimizes the hardware itself.

\subsection{ {Motivation and state-of-the-art comparison}}

Table~\ref{t1} presents a comparison of state-of-the-art frameworks for IMC hardware optimization for neural networks, emphasizing hardware parameters and hardware search space characteristics.  {In contrast to single-workload optimization, this work proposes a joint hardware-workload co-optimization approach, aiming to identify the most efficient IMC system parameters capable of supporting diverse workloads through design space exploration} (Fig.~\ref{fmotive}). 
Most existing frameworks focus on optimizing crossbar size and the number of bits per IMC device (relevant primarily for non-volatile memory technologies, like RRAM) \cite{negi2022nax}, as these are core processing elements in IMC architectures.  {The hardware search beyond the crossbar macro is relatively limited, and architectural and system-level parameters are less commonly explored.} In this work, we address the limited hardware parameter exploration in state-of-the-art methods by targeting hardware parameter optimization across multiple levels of IMC hardware design hierarchy, including device-, circuit-, architecture-, and system-level parameters,   {accounting for interdependencies among them. In addition, we target optimization across a broader range of supported workloads and improve the optimization algorithm (see Section} \ref{Salg}).

\begin{figure*}[ht]
    \includegraphics[width=\textwidth]{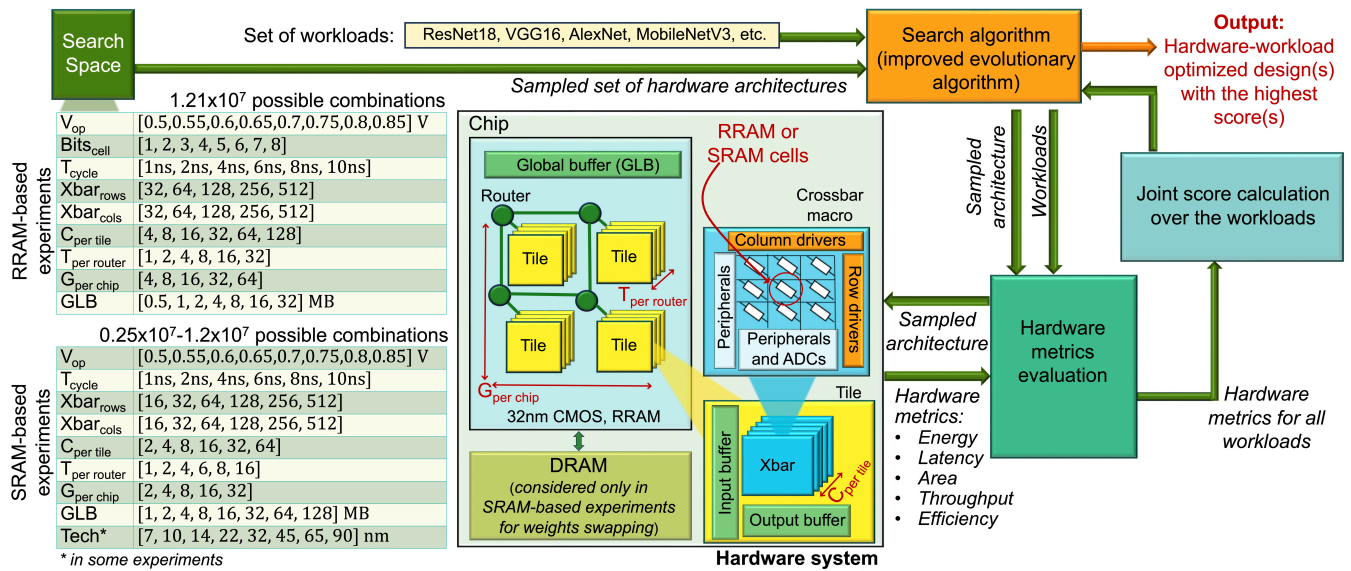}
    \caption{Proposed joint hardware-workload co-optimization framework for in-memory computing hardware.}
    \label{ffr}
\end{figure*}

\section{Proposed joint hardware-workload co-optimization framework}
\label{Sproposed}

\subsection{Framework overview}

The proposed joint hardware-workload co-optimization framework for IMC-based hardware design is illustrated in Fig.~\ref{ffr}. The primary objective of this approach is to perform hardware design space exploration and identify optimal hardware parameters for a generalized IMC-based system that can efficiently support a diverse set of workloads. The framework consists of a defined hardware search space, an optimization algorithm, methods for hardware metric evaluation, and a scoring mechanism for joint performance assessment. The primary inputs to the framework are a set of target workloads, such as neural network models to be supported by the hardware, and the hardware parameter search space defined for the IMC-based architecture. The search space includes the key hardware parameters considered for optimization during the design space exploration process (Section \ref{SSpace}). 
In this work, we evaluate the core functionality of the framework on a diverse range of CNN model types, including ResNet18, VGG16, AlexNet, and MobileNetV3, which differ significantly in size, structure, and their corresponding mapping characteristics to the hardware. Then, we further extend the framework to additional workloads, including MobileBERT, DenseNet201, ResNet50, Vision Transformer (ViT), and GPT-2 Medium, thereby demonstrating its adaptability and generalization capability.

The sampled hardware parameters from the search space, along with the set of target workloads, are provided as inputs to the optimization algorithm. In this work, we adopt and modify an evolutionary algorithm, as discussed in Section~\ref{Salg}, as it is well-suited and efficient for exploring discrete search spaces. The optimization process relies on a joint score calculated across all workloads, incorporating multiple hardware metrics, such as energy consumption, latency, and on-chip area, depending on the selected objective function. In contrast to many state-of-the-art frameworks that focus on a single workload (e.g., a specific neural network model) and optimize hardware parameters based solely on its feedback, our framework evaluates designs based on a combined score across all workloads. This enables the selection of IMC-based hardware configurations that are optimized and generalized for diverse workloads. The hardware evaluation is performed using the CIMLoop framework~\cite{andrulis2024cimloop}, which offers fast and flexible support for diverse hardware configurations. It is based on real IMC crossbar data and provides accuracy comparable to analytical, cycle-based simulators such as NeuroSim~\cite{peng2020dnn+}, while significantly reducing simulation time, which is a critical advantage for large-scale hardware exploration experiments. 
Even though such frameworks do not achieve the accuracy of silicon validation or detailed circuit-level simulators, they provide sufficiently reliable estimates for relative design comparison and offer fast evaluation, which is essential for the large search space considered in this work.

\subsection{Search space and hardware configuration setups}
\label{SSpace}

In this work, we target tiled in-memory computing architectures that leverage hierarchical organization across crossbars, tiles, and system-level components. In contrast to the majority of hardware exploration frameworks, which primarily target the crossbar macro, a relatively small component within the overall IMC system~\cite{negi2022nax, yang2021multi, sun2023gibbon, moitra2023xpert}, we emphasize that optimizing all levels of the system hierarchy is essential for achieving hardware-efficient IMC chip designs. 

We conduct experiments on two distinct types of IMC hardware: RRAM-based and SRAM-based architectures, which differ fundamentally in how neural network weights are stored and processed.
For RRAM-based designs, we adopt a purely weight-stationary approach and assume that all weights must fit on-chip. In contrast, for SRAM-based designs, we allow weight swapping, such that only a subset of layers' weights are stored on-chip at a time. Once these layers are processed, the weights are swapped out and replaced with those for the next layers. This approach is more appropriate for SRAM-based IMC architectures, as SRAM offers faster write times and better endurance compared to RRAM. Moreover, due to the larger area overhead of SRAM, it is often impractical to fit all model weights within the area constraints of a single chip, particularly for larger networks.
By evaluating these two fundamentally different memory architectures and system-level constraints, we are able to assess and compare hardware design strategies under diverse and realistic scenarios. These experiments also demonstrate that our framework is versatile and beneficial across different memory technologies and design assumptions, making it well-suited for a wide range of IMC hardware exploration use cases.

The tiled IMC architecture used in this work is illustrated in Fig.\ref{ffr}. For both scenarios, we simulate a tiled, crossbar-based architecture incorporating different memory technologies, RRAM- or SRAM-based, using device models from \cite{lu2021neurosim}. Each crossbar macro comprises memory cells along with peripheral circuits, including column and row drivers, input/output buffers, and analog-to-digital converters (ADCs) (modeled from ~\cite{lu2021neurosim}).  {The input to the crossbar is modeled as a 1-bit activation bit-stream, applied sequentially across the array. The design assumes one ADC per crossbar macro, and does not explore optimization of the number of ADCs or column sharing among multiple columns, since ADCs already constitute a substantial overhead at the macro level}~\cite{krestinskaya2022towards}. Multiple crossbar macros and their associated peripheral components, e.g., input/output buffers, form a single tile. Several tiles share a router for data transmission, following the architecture proposed in~\cite{shafiee2016isaac}. At the chip level, the system consists of groups of tiles, interconnect routers, and a global buffer. For SRAM-based experiments involving weight swapping, an external DRAM component is included. Specifically, we utilize LPDDR4 to support off-chip weight storage and loading. LPDDR4 is selected due to its low power consumption and high bandwidth \cite{he2022design, yu2021compute}.

The search space size ranges from $0.25 \times 10^7$ to $1.21 \times 10^7$ possible combinations of hardware parameters (depending on the experiment), including those illustrated in Fig.\ref{ffr} and summarized in Table \ref{t1}. The search space includes crossbar dimensions (number of rows $\mathrm{Xbar_{rows}}$ and columns $\mathrm{Xbar_{cols}}$), number of crossbars per tile ($\mathrm{C_{per\,tile}}$), number of tiles per router ($\mathrm{T_{per\,router}}$), and the number of tile groups per chip ($\mathrm{G_{per\,chip}}$). The crossbar within a tile refers to the crossbar macro, which includes both the crossbar array itself and the associated peripheral circuits, such as buffers, ADCs, and related components. Additionally, the operating voltage ($\mathrm{V_{op}}$), which affects both energy efficiency and throughput~\cite{andrulis2024cimloop}, is included as an optimization variable. The cycle time ($\mathrm{T_{cycle}}$), representing the operating frequency, and the size of the global buffer ($\mathrm{GLB}$), which stores input and output data, are also considered.
For the RRAM-based experiments, the number of bits per memory cell ($\mathrm{Bits_{cell}}$) is included as an optimization parameter, as it directly impacts the hardware configuration and determines the number of devices required to represent the weights. The majority of the experiments are conducted using the 32nm CMOS technology node. However, in the experiments presented in Section~\ref{Stradeoffs}, the CMOS technology node is also included as a variable, with the operating voltage range adjusted accordingly.
In weight-stationary RRAM-based designs, the search space includes slightly larger values for $\mathrm{Xbar_{rows}}$, $\mathrm{Xbar_{cols}}$, $\mathrm{C_{per\,tile}}$, $\mathrm{T_{per\,router}}$, and $\mathrm{G_{per\,chip}}$, as all network weights are required to fit on-chip. In contrast, for SRAM-based designs utilizing weight swapping, a wider range of $\mathrm{GLB}$ sizes is explored to accommodate not only the intermediate input/output data but also the temporary storage of weights during layer swapping.

 {In this work, we focus on architecture-level modeling of energy, latency, and area, and therefore do not explicitly model thermal gradients or long-term device aging effects. Accurate thermal and reliability analysis would require layout- and time-dependent models and technology-calibrated parameters, substantially increasing design-space complexity and search time, and is thus beyond the scope of this work.}

\subsection{Optimization algorithm}
\label{Salg}

\subsubsection{ {Why evolutionary genetic search?}}

 As per Table~\ref{t1}, most of the state-of-the-art works use evolutionary algorithms (EA), Bayesian optimization (BO), or differential search (DS) methods, while some optimization frameworks for neural network parameters with hardware feedback  additionally rely on reinforcement-learning-based (RL) techniques \cite{jiang2020device}. { In this work, the hardware optimization problem involves a large-scale, high-dimensional, and predominantly discrete design space. The search space spans architectural, circuit-level, memory-organization, and technology-dependent parameters, many of which are integer-valued, categorical, or conditionally dependent and subject to hard architectural and workload-specific constraints. These characteristics result in a non-convex and non-smooth objective landscape with frequent feasibility discontinuities.}

 {Under such conditions, alternative optimization strategies (non-EA) are poorly matched to the problem structure. BO, DS, and RL require significant modeling assumptions, surrogate construction, or training overhead. BO struggles with high-dimensional categorical variables and conditional constraints, DS methods rely on differentiability assumptions that are violated by discrete hardware parameters, and RL requires training a controller over a very large combinatorial space with discontinuous, constraint-driven rewards and high evaluation cost} \cite{krestinskaya2025cimnas}. 
 {In contrast, EA operates directly on the native discrete parameterization, enables efficient population-based exploration and pruning of infeasible designs, and avoids surrogate or controller training. Therefore, we adopt an EA approach for this work. The comparison between the methods is summarized in Table} \ref{tab:opt_method_comparison}.

\begin{table}[ht]
\centering
\caption{ {Comparison of optimization methods for discrete hardware design-space exploration.
$\checkmark$: well suited, $\triangle$: possible with overhead, $\times$: poorly suited.}}
\resizebox{\columnwidth}{!}{%
\label{tab:opt_method_comparison}
\begin{tabular}{lcccc}
 \hline
 {\textbf{Aspect}} &  {\textbf{EA}} &  {\textbf{RL}} &  {\textbf{BO}} &  {\textbf{DS}} \\
 \hline
 {Discrete and categorical variables} &  {$\checkmark$} &  {$\triangle$} &  {$\triangle$} &  {$\triangle$} \\
 {Large combinatorial space ($10^6$--$10^7$)} &  {$\checkmark$} &  {$\triangle$} &  {$\times$} &  {$\times$} \\
 {Hard and conditional constraints} &  {$\checkmark$} &  {$\triangle$} &  {$\triangle$} &  {$\times$} \\
 {Non-smooth objectives} &  {$\checkmark$} &  {$\triangle$} &  {$\times$} &  {$\times$} \\
 {Expensive hardware evaluation} &  {$\checkmark$} &  {$\times$} &  {$\times$} &  {$\times$} \\
 {Extra modeling or training required} &  {$\times$} &  {$\checkmark$} &  {$\checkmark$} &  {$\checkmark$} \\
 \hline
\end{tabular}
}
\end{table}

 {To select the starting algorithm among evolutionary and evolutionary-inspired approaches, we evaluated several optimization methods on a reduced RRAM-based search space defined by $\mathrm{Xbar_{rows}}$, $\mathrm{Xbar_{cols}}$,  $\mathrm{C_{per\,tile}}$, and $\mathrm{Bits_{cell}}$, while fixing the remaining parameters. All architectures within this reduced space were first exhaustively evaluated in terms of their hardware metrics, allowing the identification of both local and global minima. We then compared the performance of Genetic Algorithm (GA), Particle Swarm Optimization (PSO)}~\cite{kennedy1995particle},  {Evolutionary Strategy (ES), Stochastic Ranking Evolutionary Strategy (ERES)}~\cite{runarsson2000stochastic},  {Covariance Matrix Adaptation Evolutionary Strategy (CMA-ES)}~\cite{runarsson2000stochastic}, { and G3PCX}~\cite{deb2002computationally}.  { The results of these experiments are summarized in Table}~\ref{tab:alg_speed_global}.  { Among the evaluated methods, only GA, ES, and ERES consistently converged to global minima. PSO and G3PCX converged to optimized designs but were trapped in local minima, while CMA-ES did not converge for the considered optimization problem. Among the algorithms that reached global minima, GA exhibited the fastest convergence in terms of search time (about 1.5$\times$ faster than ES/ERES). Therefore, GA was selected as the baseline optimization algorithm for this work. The proposed modifications to the GA are discussed in Section}~\ref{subsecproposedAlg}.

\begin{table}[ht]
\centering
\caption{ { Comparison of evolutionary and genetic-inspired optimization algorithms on a small search space.}}
\label{tab:alg_speed_global}
\begin{tabular}{lcc}
 \hline
 { \textbf{Algorithm}} &  {\textbf{Global minima}} &  {\textbf{Relative speed}} \\
 \hline
 { GA} &  { $\checkmark$} &  { Fastest} \\
 { ES/ERES} &  { $\checkmark$} &  { Moderate} \\
 { PSO} &  { $\times$ (local minima)} &  { Fast} \\
 { G3PCX} &  { $\times$ (local minima)} &  { Moderate} \\
 { CMA-ES} &  { $\times$ (no convergence)} &  { Slow} \\
 \hline
\end{tabular}
\end{table}

\subsubsection{ {Proposed optimization algorithm}}
\label{subsecproposedAlg}

Algorithm \ref{alg} outlines the main steps of the joint hardware-workload co-optimization process, incorporating the proposed algorithm with modified sampling and a four-phase optimization strategy. We observed that using purely random sampling for initializing the population in the evolutionary algorithm significantly influences the final design selection and convergence behavior. To address this, we modified the sampling procedure to improve convergence and ensure experimental repeatability regardless of the initial random seed.
The revised sampling method for the initial population consists of three steps. First, a preliminary set of $P_H$ hardware candidates, denoted as $C_1$ $=$ $\{c_{1-1}, c_{1-2}, ..., c_{1-P_H}\}$, is randomly sampled from the hardware design space. In this work, we set $P_H$ $=$ $1000$, as this step is not computationally intensive. If weight swapping is not employed and the entire neural network model must fit on the hardware (e.g., in RRAM-based designs), we further constrain the initial population to only include designs that can accommodate the largest workload.
Next, we enhance diversity by selecting the most distinct candidates based on Hamming distance \cite{labib2019hamming}, thereby promoting broader exploration of the design space. This step ensures that the optimization process begins with a diverse set of candidates, increasing the likelihood of identifying high-quality solutions.
The Hamming distance \( d_H(X, Y) \) between two design samples \( X \) and \( Y \), each characterized by  \( n \) discrete parameters, is defined as \cite{labib2019hamming}:
\begin{equation}
d_H(X, Y) = \sum_{i=1}^{n} 1(X_i \neq Y_i), 
\end{equation}
where  \( X_i, Y_i \) denote the \( i \)-th parameters of designs \( X \) and \( Y \), respectively. The indicator function $1(X_i \neq Y_i)$  returns 1 when the two parameters differ and 0 otherwise \cite{labib2019hamming}:
\begin{equation}
    1(X_i \neq Y_i) =
    \begin{cases}
        1, & \text{if } X_i \neq Y_i \\
        0, & \text{otherwise}
    \end{cases}
\end{equation}

To construct the set of the most distinct designs,  $C_2=\{c_{2-1}, c_{2-2},...,c_{2-P_E}\}$, we employ a greedy selection algorithm that adds one design at a time.  The set $C_2$ is initialized with the first candidate from  $C_1$, e.g. $C_2=\{c_{1-1}\}$. Then, until $C_2$ contains  $P_E$ designs (here $P_E=500$), we iteratively select one design from the remaining candidates in $C_1$ that is most dissimilar to those already in $C_2$. For each unselected candidate design $X$,  we compute its minimum Hamming distance to the current set $C_2$, defined as $d_{\min}(X, C_{2}) = \min_{Y \in C_{2}} d_H(X, Y)$. The candidate with the highest $d_{\min}$ value,  e.g. the one farthest from its closest counterpart in $C_2$, is then added to $C_2$. This strategy ensures that the resulting set $C_2$ consists of the most diverse hardware designs for further evaluation.

\begin{algorithm}[t!]
    \caption{Joint hardware-workload co-optimization algorithm with EDAP-based objective function.}\label{alg}
    \begin{algorithmic}
        \STATE  $\mathbf{Inputs:}$
        \STATE  $\; \mathrm{Set \, of \, workloads: \,}  W = [w_1, w_2,..., w_n]$
        \STATE  $ \; \mathrm{Hardware \,search \,space: \,} S = [\mathrm{V_{op}, Bits_{cell}, ..., GLB}]$
        \STATE  $\mathit{Initial \, population \, sampling:}$
        \IF{$\mathrm{Weights \, swapping \, case}$}
             \STATE $\mathrm{Randomly \,\mathbf{sample} \,} P_H \mathrm{\, candidate \,hardware}$
             \STATE  $ \quad \mathrm{\,configurations \,from\, search \, space \,} S $
       \ELSE
             \STATE  $\mathbf{Find} \, M \, \mathrm{memory \, \,elements \,required \, for\, the\, largest\,} w_i$
            \STATE  $ \qquad \mathrm{\,workload} $
            \WHILE{$(\mathrm{population} \, p < P_H)$}
                \STATE $\mathrm{Randomly \,\mathbf{sample} \,candidate \,hardware \, }  c_h$ 
                \IF{$\mathrm{Number \, of  \,memory \, elements \, of \, } c_h \, \geq M$}
                    \STATE $\mathrm{\mathbf{Keep} \, the \, sample \, in\, the \,population}$
                    \STATE $p=p+1$
                \ENDIF
           \ENDWHILE
       \ENDIF
       \STATE  $\mathbf{Select} \, P_E \, \mathrm{candidates \,based \,on \,the \,  highest\,} d_{H}$ 
        \FOR{$\mathrm{each \, candidate \,} \alpha \mathrm{\, in \,} C_2$}
        \STATE $\mathrm{Find  \,area \,} A$
          \FOR{$w_i \, \mathrm{in} \, W$}
            \STATE $\mathrm{\mathbf{Evaluate}  \,energy \,} E_{wi} \, \mathrm{and \, latency} \, L_{wi} $
        \ENDFOR
        \STATE $\mathrm{\mathbf{Calculate} \,joint\, score\,} f_\alpha=\max(E_w)\times \max(L_w) \times A$
        \ENDFOR
        \STATE $\mathrm{\mathbf{Select} \,} P_{GA} \mathrm{\, candidates \, with \, lowest \, scores}$
        \STATE
        \STATE  $\mathit{Four-phase \, search:}$
        \FOR{$\mathrm{phase} \, \mathbf{in} \, \mathrm{[Exploration, Transition, Convergence,} \newline \mathrm{Finetuning] \, with \, corresponding \,} \mathbb{P}_c , \eta_c, \mathbb{P}_m,  \eta_m \mathrm{\,values}$}
        
        \FOR{$ G \,\mathrm{generations}$}
            \STATE  $\mathit{Evaluation:}$
            \FOR{$\mathrm{each \,sample} \, \alpha \, \mathrm{in \, a \, population} $}
                \STATE $\mathrm{\mathbf{Evaluate} \, hardware \, metrics}$
                \STATE $\mathrm{\mathbf{Calculate} \, score\,} f_{\alpha} \,\mathrm{(same \,as \,in \,sampling)} $
            \ENDFOR
            \STATE  $\mathit{Selection, \, crossover, \, and\, mutation:}$
            \STATE $\mathrm{\mathbf{Sort}\, the  \,designs }$
            \STATE $\mathrm{\mathbf{Select} \, designs \, to \, participate\,in \, crossover \,}$
            \STATE $ \quad \mathrm{ (crossover\, probability\,} \mathbb{P}_c \mathrm{)}$
            \STATE $\mathrm{\mathbf{Perform} \,crossover \,with  \,distribution \, \eta_c, \,}$
            \STATE $\quad \mathrm{ constructing\, new\,  candidates}$
            \STATE $\mathrm{\mathbf{Execute} \,mutation \, of\, new\, candidates \,with }  $
            \STATE $ \quad \mathrm{probability \,} \mathbb{P}_m \mathrm{\, and \,} \mathrm{distribution \, \eta_m, \, constructing \,  }$
            \STATE $ \quad \mathrm{new \, population \,with \,} P_{GA} \mathrm{\, samples}$
        \ENDFOR
        \ENDFOR
    \end{algorithmic}
\end{algorithm}

After selecting the set of diverse candidates $C_2$, we evaluate the performance of each design based on a combined score across all workloads. For instance, in an optimization problem targeting the Energy-Delay-Area Product (EDAP), the performance score can be calculated as:
\begin{equation}\label{eq:o1}
f(E_w, L_w, A)=\max(E_w)\times\max(L_w)\times A,
\end{equation}
where $E_w=\{E_{w1}, E_{w2}, ... E_{wn}\}$ represents the energy consumption for processing $n$ different workloads (e.g., corresponding to different neural network models), $L_w=\{L_{w1}, L_{w2}, ... L_{wn}\}$ denotes the associated latencies, and $A$ is the on-chip area of the design candidate.
Once the performance of all candidates in $C_2$ is evaluated, we construct the final initial population for the genetic algorithm by selecting the top $P_{GA}$ candidates with the lowest scores.
In this study, the initial population is not sampled randomly, but is instead composed of well-performing and diverse designs. This targeted selection narrows the search space toward promising regions, leading to improved convergence toward the global minimum and avoiding suboptimal local minima. Furthermore, this approach ensures greater repeatability of experiments compared to traditional Genetic Algorithms (GAs), where final outcomes are often highly sensitive to the randomness of the initial sampling process.

After sampling the initial population, the four-phase Genetic Algorithm (GA) is executed. The four phases, exploration, transition, convergence, and fine-tuning, differ in their crossover and mutation parameters, as summarized in Table~\ref{t2}. Each phase runs for $G$ generations. In this study, we use the same value of 
$G$ for all phases to maintain uniformity in phase duration. The optimization problem for the Genetic Algorithm is formulated as  $f (E_w, L_w, A)$, $\mathrm{s.t.} \, A \leq A_{constr}$.

In this work, we focus on area-constrained optimization with an area constraint $A_{constr}$, ensuring that the optimized designs remain within practical limits for implementation on a single chip, particularly important for RRAM-based architectures. One example of a possible objective function is shown in Eq.\ref{eq:o1}. Additional experiments using alternative objective functions are presented in Section \ref{results}.
Each phase of the algorithm begins with an evaluation step, where hardware simulation is conducted for every design in the population, corresponding hardware metrics are computed, and a performance score is assigned to each sample. This is followed by the crossover operation, in which new architectures are generated by recombining parameters from designs in the previous generation. Lastly, a mutation step is applied, introducing variations into the newly formed architectures to promote diversity and further exploration of the search space.

\begin{table}[t!]
\centering
\caption{Parameters for different search phases.}
\resizebox{\columnwidth}{!}{%
\begin{tabular}{|l|c|c|c|c|}
\hline
\textbf{Phase} & \textbf{Crossover $\mathbb{P}_c$} & \textbf{Crossover $\eta_c$} & \textbf{Mutation $\mathbb{P}_m$} & \textbf{Mutation $ \eta_m$} \\ \hline
Exploration    & 1.0                  & 3.0                  & 1.0                 & 3.0                 \\ \hline
Transition     & 0.9                & 7.0                  & 0.5               & 7.0                 \\ \hline
Convergence    & 1.0                  & 15.0                 & 0.2               & 15.0               \\ \hline
Fine-tuning    & 1.0                  & 25.0                 & 0.05              & 25.0                \\ \hline
\end{tabular}
}
\label{t2}
\end{table}

This work uses simulated binary crossover and polynomial mutation operators~\cite{deb2007self, blank2020pymoo}, with phase-specific parameters summarized in Table~\ref{t2}. The crossover and mutation probabilities, $\mathbb{P}_c$ and $\mathbb{P}_m$, range from 0 to 1, while the distribution indices, $\eta_c$ and $\eta_m$, typically range from 3 to 30. Probabilities determine the fraction of candidates undergoing crossover or mutation, whereas distribution indices control the extent of variation, particularly since lower values yield larger changes and higher values produce finer adjustments.
In the exploration phase, all candidates undergo crossover ($\mathbb{P}_c = 1.0$, $\eta_c = 3$) and mutation ($\mathbb{P}_m = 1.0$, $\eta_m = 3$), allowing offspring to differ significantly from their parents and ensuring strong diversity and broad coverage of the design space, which helps avoid convergence to local minima.
In the transition phase, diversity is preserved while refining promising designs, with $\mathbb{P}_c = 0.9$ (enabling some direct inheritance), $\eta_c = 7$ (maintaining variation), and $\mathbb{P}_m = 0.5$, balancing exploration and exploitation.
In the convergence phase, the search focuses on high-performing regions using higher distribution indices ($\eta_c = 15$, $\eta_m = 15$) to generate small focused variations and a reduced mutation probability ($\mathbb{P}_m = 0.2$) to preserve high-quality solutions.
Finally, the fine-tuning phase applies precise adjustments, with very high distribution indices ($\eta_c = 25$, $\eta_m = 25$) and minimal mutation ($\mathbb{P}_m = 0.05$), keeping offspring close to their parents while refining top solutions.

\section{Results}
\label{results}

The primary hardware simulations are conducted using the 32nm CMOS technology node, except for the experiments in Section~\ref{Stradeoffs}, where the technology node is included as a variable in the optimization process. All neural network models used in the experiments are quantized to 8 bits for both weights and activations to enable efficient hardware mapping.
In most experiments, the number of generations for all optimization algorithms is fixed at $G = 10$, and the population size of the genetic algorithm during the evaluation phase (following the sampling phase) is set to $P_{\mathrm{GA}} = 40$. An exception is made in the trade-off analysis experiments, where a larger population size of $P_{\mathrm{GA}} = 70$ is used to better explore the broader parameter space. To ensure that the resulting chip designs remain within practical fabrication constraints for large-die accelerators~\cite{nvidia_tensor_core}, an area constraint of $A \leq 800\mathrm{mm}^2$ is applied across all simulations.
It is also important to note that, in SRAM-based experiments, off-chip DRAM is excluded from the on-chip area calculation, as it does not contribute to the physical footprint of the fabricated chip. However, the energy consumption and latency associated with data transfers between DRAM and on-chip components are fully considered in the evaluation.
In this section, we refer to the joint hardware-workload  co-optimization approach as 'joint search' or 'joint optimization' and to optimization over individual workloads separately as 'separate search' or 'separate optimization'. 

\begin{figure}[t]
    \centering
    \includegraphics[width=\columnwidth]{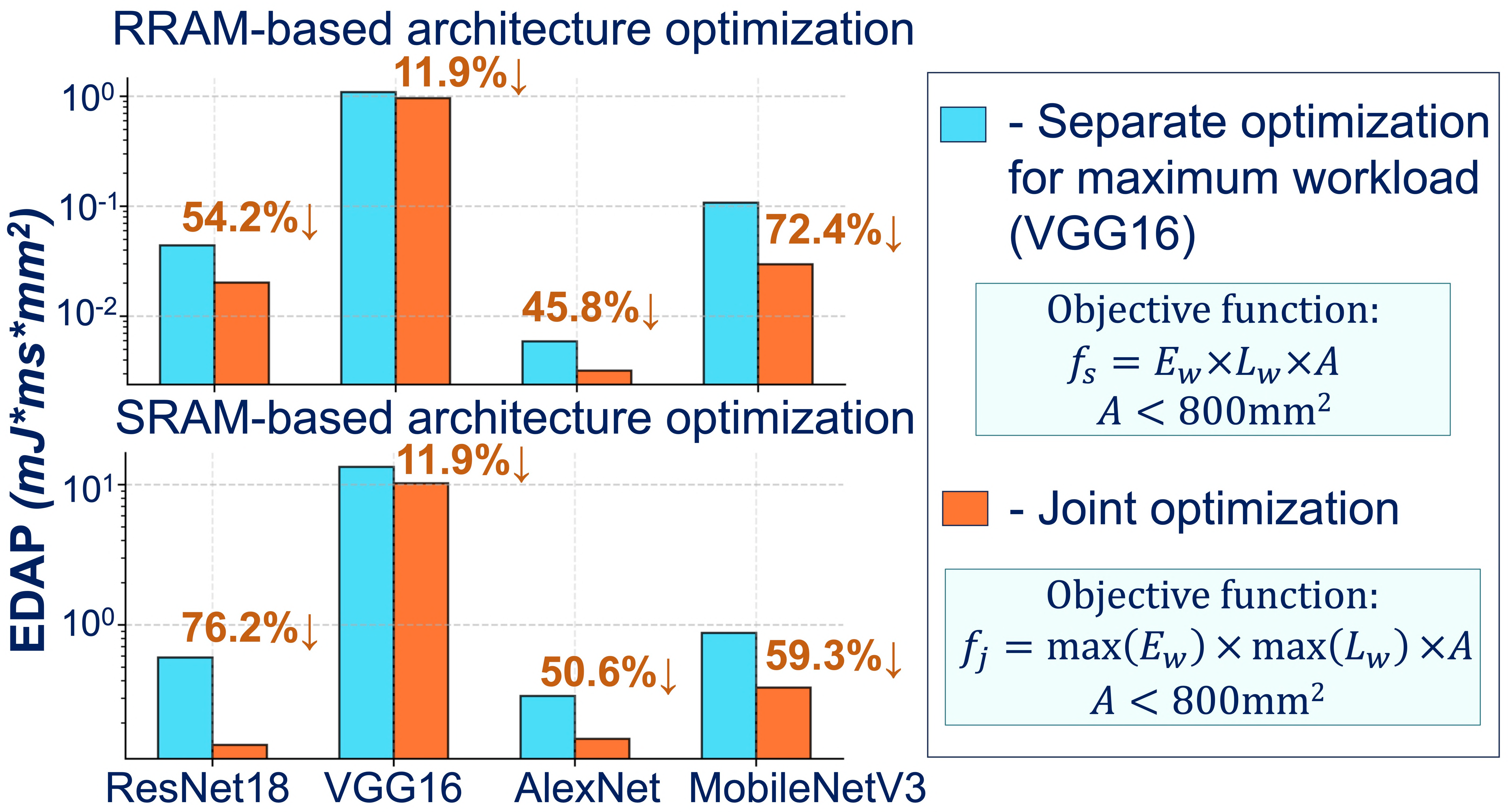}
    \caption{EDAP comparison of optimized designs for RRAM-based and SRAM-based IMC hardware, obtained using separate optimization for the largest workload and joint optimization across multiple workloads.}
    \label{f1p2}
\end{figure}

\subsection{Comparison of joint hardware-workload co-optimization with optimization for the largest workload}
\label{SRA}

First, we compare the proposed approach with optimization targeting only the maximum workload, e.g., the neural network model with the largest size among all target workloads to be supported by the chip (VGG16 in our case).
Optimization for the maximum workload represents a naive approach commonly used in hardware design, where the system is optimized only for the most demanding model. Although optimizing for the largest workload ensures coverage of the most demanding case, we show that it often compromises efficiency when applied to smaller or structurally different models.
The comparison is demonstrated in Fig.~\ref{f1p2}, which presents the actual EDAP scores for four different workloads using the top-1 designs obtained from each approach.
Results are shown for both RRAM-based and SRAM-based hardware scenarios. Across all networks, the joint optimization approach consistently achieves lower EDAP scores compared to optimization based solely on the largest workload. Notably, even for the largest workload, the EDAP score resulting from joint optimization is lower than that obtained from the equivalent single-workload optimization. This can be attributed to the influence of smaller workloads in the joint optimization process, which drives the algorithm to more aggressively minimize EDAP across the entire workload set, often leading to better overall designs within the same number of generations and population size constraints.

\subsection{Performance of the proposed algorithm: convergence improvements and the impact of optimized sampling}
\label{SRB}

\begin{figure}[t]
    \includegraphics[width=\columnwidth]{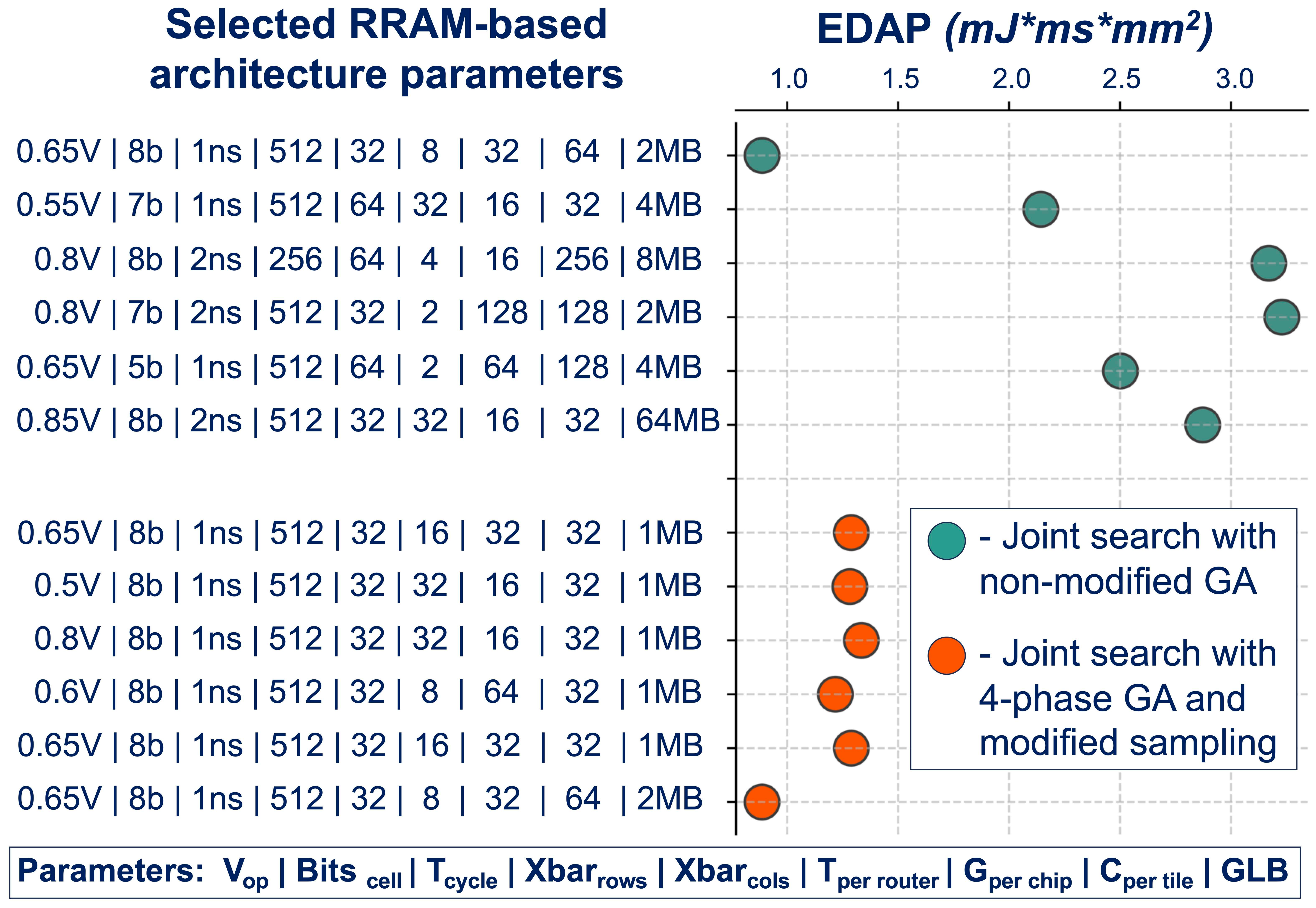}
    \caption{Improved convergence and EDAP scores of optimized designs achieved using the proposed 4-phase genetic algorithm (GA) with enhanced sampling, compared to the traditional, non-modified GA approach (six independent experiments with RRAM-based hardware for each case).}
    \label{f2}
\end{figure}

Even though evolutionary algorithms are well-suited for discrete search spaces in hardware design space exploration~\cite{krestinskaya2024towards, sun2023gibbon, krestinskaya2020towards, krestinskaya2020automating}, we observed that both the convergence of the algorithm and the quality of the optimized hardware designs are highly sensitive to the initial sampling.
Fig.~\ref{f2} compares the performance of the traditional non-modified genetic algorithm (GA), as presented in \cite{krestinskaya2024towards}, with the proposed four-phase GA with enhanced sampling. The figure presents results from six independent joint optimization runs for each approach, with randomly sampled initial populations, using the EDAP objective function (same as in Fig.~\ref{f1p2} and Section~\ref{SRA}). The traditional non-modified GA shows significant variability across runs, often converging to different local optima due to its sensitivity to the initial population.
In contrast, the proposed 4-phase GA uses a more robust sampling strategy: designs are initially selected based on Hamming distance scores, followed by evaluation of candidates with the highest diversity. In addition, it performs GA in 4 separate phases with different $\mathbb{P}_c$, $\mathbb{P}_m$, $\eta_c$, and $\eta_m$ values (described in Section \ref{Salg}). This results in better convergence behavior, consistently lower EDAP scores, and improved stability and consistency of the results across independent experiments. 
In addition, to verify the consistency of our findings, we repeated each experiment for 25 additional independent runs. The simulation outcomes showed the same trend, where the 4-phase GA consistently outperformed the non-modified GA. The mean and standard deviation ($std$) of the EDAP values were $2.47$ mJ$\cdotp$ms$\cdotp$mm$^2$ ($std=0.87$) for the non-modified GA and $1.21$ mJ$\cdotp$ms$\cdotp$mm$^2$ ($std=0.16$) for the proposed 4-phase GA.

\subsection{ {Aggregation schemes testing}}
\label{subsec_newAggregation}

 {
To evaluate how different aggregation schemes affect the design-space search, in addition to the maximum-based aggregation defined in Eq.}~\ref{eq:o1}  {(denoted as \emph{Max}), we consider two alternative strategies. The first aggregates hardware metrics across all workloads by jointly considering all energy and latency values (denoted as \emph{All}), while the second uses a mean-based aggregation (denoted as \emph{Mean}). For the \emph{All} aggregation scheme with four workloads, the objective function is defined as $f(E_w, L_w, A)=E_{w\text{-all}}\times L_{w\text{-all}}\times A$, where the aggregated energy is computed as $E_{w\text{-all}}=E_{w1}E_{w2}E_{w3}E_{w4}$ and the aggregated latency is defined similarly as $L_{w\text{-all}}=L_{w1}L_{w2}L_{w3}L_{w4}$. For the \emph{Mean} aggregation scheme, the objective function is defined as $f(E_w, L_w, A)=\text{mean}(E_w)\times\text{mean}(L_w)\times A$. The simulation results for both RRAM- and SRAM-based designs, in terms of EDAP and total search time, are summarized in Table}~\ref{tab:rram_sram_edap_time},  {reporting the optimized designs obtained using the three aggregation strategies. While all schemes yield comparable optimization quality, the \emph{Max}-based aggregation consistently results in lower search time and produces designs with lower EDAP in the majority of evaluated cases.}

\begin{table}[ht]
\caption{ {EDAP per optimized designs and search time for RRAM and SRAM across different aggregation strategies.}}
\label{tab:rram_sram_edap_time}
\centering
\resizebox{\columnwidth}{!}{%

\begin{tabular}{|l|c|c|c|c|c|}

 \hline
 {Case} & \multicolumn{4}{c|}{ {EDAP ($mJ\times ms \times mm^2$)}}  &  {Search time} \\
\cline{2-5}
 &
 {ResNet18} &
 {VGG16} &
 {AlexNet} &
 {MobileNetV3} &
 {($hours$)} \\
 \hline
\multicolumn{6}{|c|}{ {\textbf{RRAM}}} \\
 \hline
 {\emph{All}}  &  {0.02051} &  {\textbf{0.88869}} &  {0.00324} &  {0.03040} &  {22.79} \\
 \hline
 {\emph{Max}}  &  {\textbf{0.02024}} &  {0.96181} &  {\textbf{0.00320}} &  {0.02974} &  {\textbf{19.89}} \\
 \hline
 {\emph{Mean}} &  {0.02040} &  {0.90902} &  {0.00322} &  {\textbf{0.02753}} &  {20.25} \\
 \hline
\multicolumn{6}{|c|}{ {\textbf{SRAM}}} \\
 \hline
 {\emph{All}}  &  {0.51744} &  {13.75740} &  {0.28962} &  {0.63487} &   {20.70} \\
 \hline
 {\emph{Max}}  &  {\textbf{0.13954}} &  {\textbf{10.22635}} &  {\textbf{0.15391}} &  {0.35698} &  {\textbf{18.13}}  \\
 \hline
 {\emph{Mean}} &  {0.15330} &  {11.59565} &  {0.15572} &  {\textbf{0.15841}} &  {19.89} \\
 \hline
\end{tabular}
}
\end{table}

\begin{figure*}[t!]
    \includegraphics[width=\textwidth]{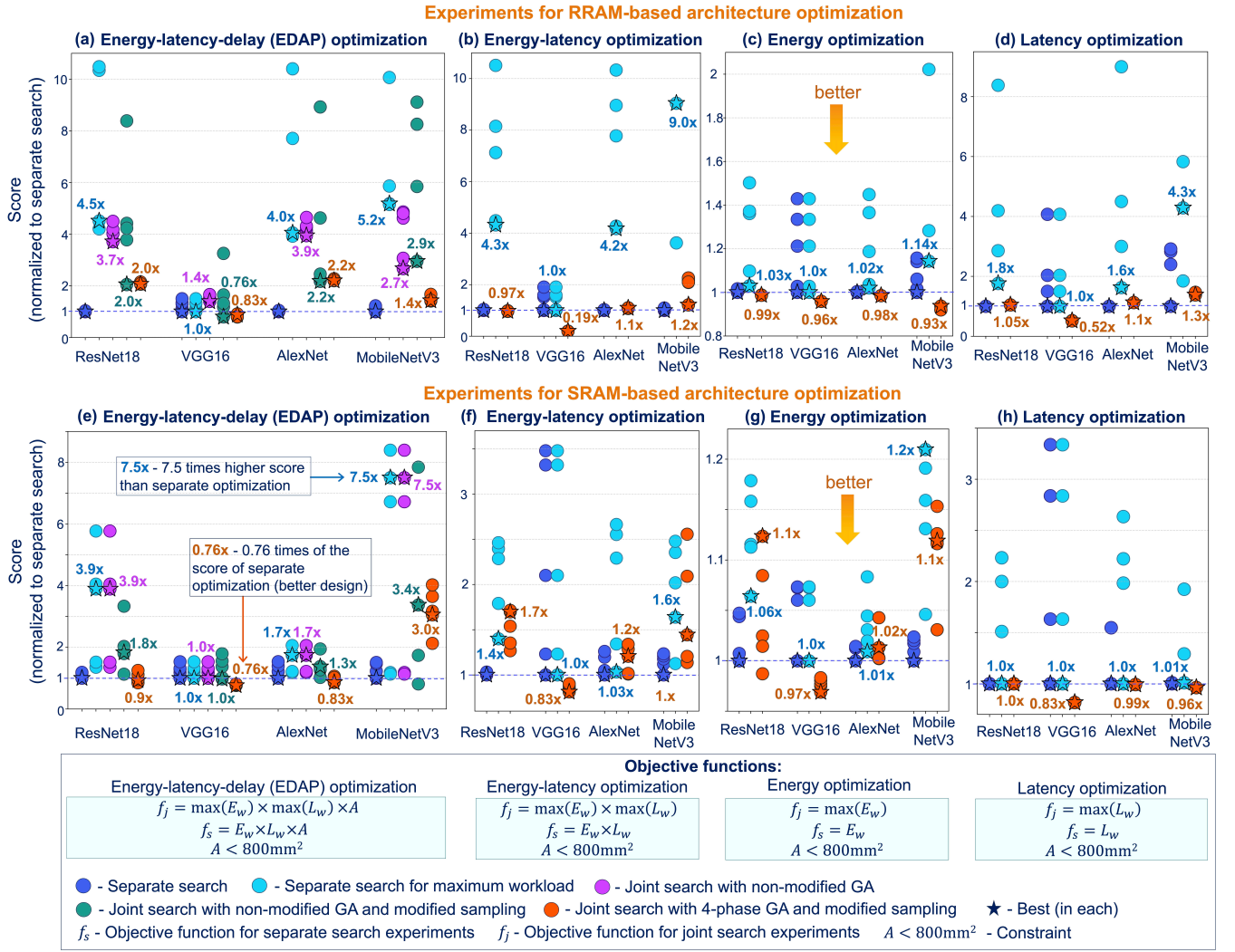}
    \caption{Comparison of optimization strategies for generating generalized architectures across  {RRAM-based (a-d) and SRAM-based (e-h)} experiments using different objective functions. These include separate search, maximum-workload-based optimization, joint search with a non-modified genetic algorithm (GA) \cite{krestinskaya2024towards}, joint search with enhanced sampling, and joint search using the proposed four-phase GA with optimized sampling  {(top-5 designs are shown in each experiment with top-1 marked with a star)}. The goal is to achieve performance that is as close as possible to the individually workload-optimized (separate search) designs, indicating minimal loss in hardware efficiency when transitioning to generalized hardware. The results demonstrate that the proposed algorithm is the most effective among all approaches in the majority of the cases, supporting the selection of optimized designs where the transition to generalized architectures results in the least compromise in hardware performance.}
    \label{f3}
\end{figure*}

\subsection{Performance gap between generalized designs and workload-specific designs}
\label{SRC}

General-purpose hardware solutions that support multiple workloads often sacrifice peak performance compared to workload-specific designs. Workload-specific designs are those that are independently optimized for each individual workload. In this work, such optimization is referred to as separate search (or separate optimization), which provides an upper bound on the achievable performance when the hardware is specifically tailored to a single workload. This work aims to reduce the performance gap between workload-specific designs and those supporting multiple workloads.
 The results for the corresponding experiments for different objective functions are shown in Fig.~\ref{f3}.
 {Each case shows the top five designs obtained from the corresponding optimization experiments, with the top-1 design marked with a star.
To ensure a fair comparison, the random seed for the initial population is set to the same value across all experiments. In addition, all experiments were repeated with five different random seeds for the initial population, yielding trends consistent with Fig.}~\ref{f3}.

In Fig.~\ref{f3},  {the optimized designs obtained via separate search are used as the baseline (dark blue), and the other experiments are normalized to this baseline (dashed blue line corresponding to the score of 1). This "separate search" score, along with its optimized designs, is derived by performing independent optimization runs for each workload. }
The scores labeled 'separate search for maximum workload' are obtained from a single optimization experiment  {per graph, where optimization is performed only for the VGG16 network as the largest workload and then evaluated on the other workloads. For all three joint search experiments, scores are obtained from a single joint optimization run per graph.} 
Then, each workload’s score is calculated using the best-selected design from that joint search. For example, if the joint objective function is defined as $\max(E_w) \times \max(L_w) \times A$, the score for ResNet18 is calculated as $E_w \times L_w \times A$, where $E_w$ and $L_w$ are the energy and latency of executing ResNet18 on the jointly optimized  {best-selected design}. All scores for a given workload are normalized to the 'separate search' baseline score for that workload.
 {In the EDAP-related optimization experiments, we compare joint searches using a non-modified genetic algorithm (GA)} \cite{krestinskaya2024towards},  {a non-modified GA with enhanced sampling, and the proposed four-phase GA. For the remaining objective functions, the non-modified GA baselines are omitted from the graphs since they follow the same trend.}

 {The proposed algorithm consistently outperforms the other baselines and achieves the lowest scores, effectively reducing the gap between hardware optimized for individual networks and generalized hardware capable of supporting multiple workloads. The only exceptions occur for energy-latency and energy-only optimization in the SRAM case} (Fig. \ref{f3}(f) and (g)),  {where the proposed method yields a slightly higher (worse) score for ResNet18 while achieving improved scores for the remaining workloads. This behavior arises from the joint optimization process, which trades off performance for one workload to better accommodate the other three. The advantage of the proposed approach over largest-workload optimization is more evident for complex objectives, for example EDAP, due to the tighter coupling between energy, delay, and area compared to single-metric optimization.}

 The additional baselines, including the non-modified GA and the non-modified GA with enhanced sampling, shown in the EDAP experiments (Fig. \ref{f3}(a) and (e)),  demonstrate that incorporating enhanced sampling ('joint search with a non-modified GA and modified sampling' in purple) yields substantially improved EDAP results. The full four-phase GA ('joint search with a 4-phase GA and modified sampling') further refines the design parameters, consistently achieving the most optimized hardware configurations. 

 In addition, multiple optimal hardware configurations with comparable performance may exist within the search space due to interdependencies among parameters across different levels of the design hierarchy. The proposed method more effectively captures this behavior, revealing a broader set of designs with identical or closely matching objective scores, as demonstrated by the top-5 designs in each experiment, which exhibit the smallest score variation compared to the baseline methods.

\subsection{ {Runtime comparison}}
\label{subruntime}

 {The runtime comparison is summarized in Table} \ref{tab:runtime_comparison},  {where the joint search is compared with the separate search under an equivalent population size and number of generations. In the separate search, the search time varies depending on the workload size for which the optimization is performed. For the proposed algorithm with modified sampling, evaluated for both RRAM and SRAM scenarios, the sampling phase introduces an overhead of approximately 30\% of the total search time. This overhead is due to repeated hardware estimation for a large number of sampled design candidates.
The separate search and the joint search with the non-modified GA also exhibit a small sampling overhead in the RRAM optimization case, since sampling is required to select only those designs in which all neural network weights can be accommodated. Although the proposed algorithm introduces additional runtime due to the sampling phase, this overhead is justified by its consistently superior performance compared to the other methods and by the improved quality of the optimized designs across both RRAM- and SRAM-based optimization scenarios. Approaches for further reducing the runtime are discussed in Section} \ref{DiscussionRuntime}.

\begin{table}[t]
\centering
\caption{ {Runtime comparison on 64 CPU cores for equivalent population size and number of generations.}}
\label{tab:runtime_comparison}
\begin{tabular}{lcc}
 \hline
 {\textbf{Method}} &  {\textbf{Search / Sampling Time}} &  {\textbf{Total Time}} \\
 \hline
 {Separate} & { Up to 0.2 h (RRAM)} &  {8.8 h--14.9 h} \\
 {Joint (non-modified)} &  {Up to 0.5 h (RRAM)} &  {12.4 h--12.6 h} \\
 {Joint (proposed)} &  {5.6 h--6.4 h} &  {18 h--19 h} \\
 \hline
\end{tabular}
\end{table}

\subsection{Design insights from optimized architectures: RRAM and SRAM comparison}
\label{SRD}

\begin{figure}[t]
    \includegraphics[width=\columnwidth]{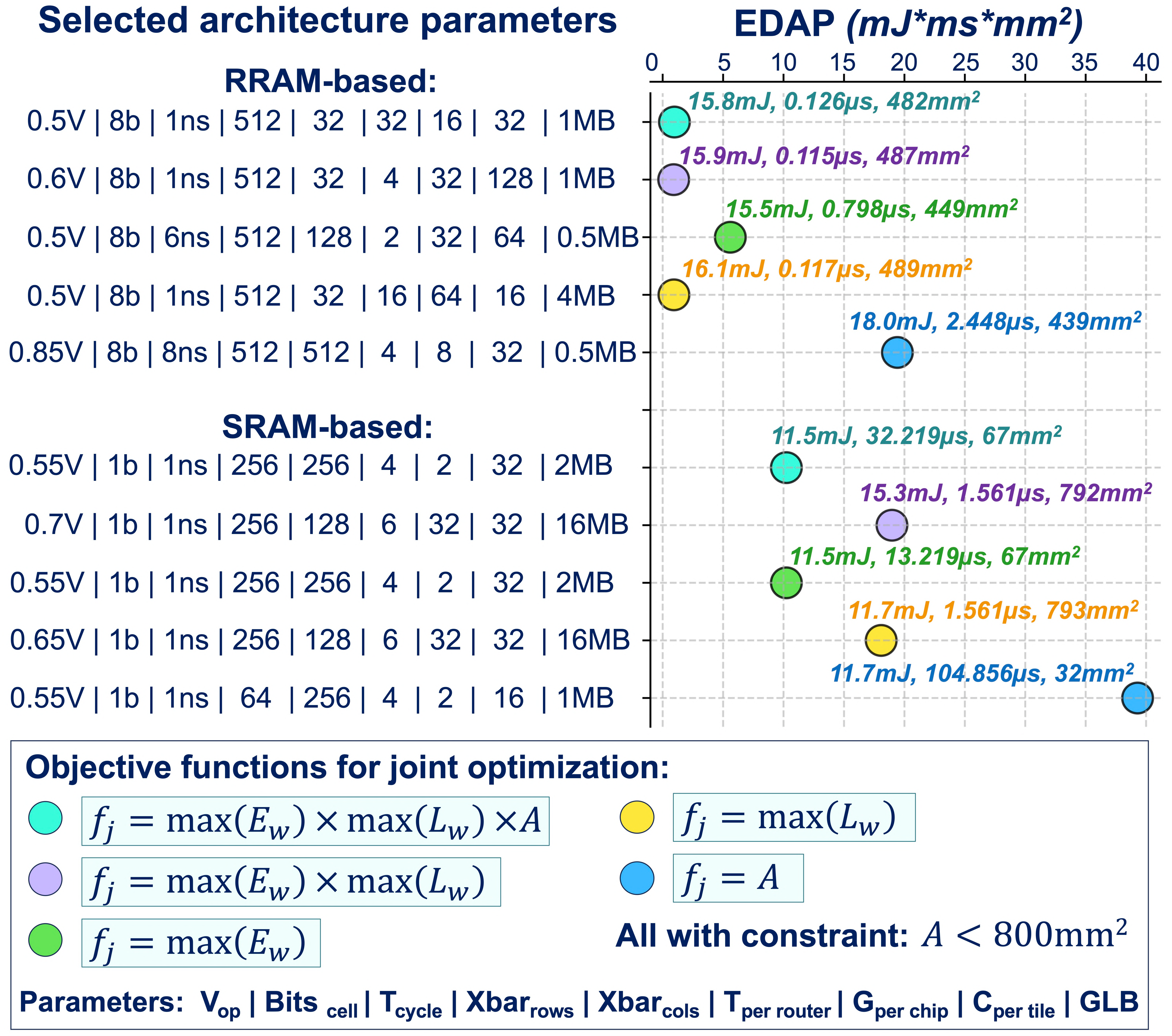}
    \caption{Comparison of optimized RRAM- and SRAM-based hardware designs across various objective functions, with energy and latency results shown for the largest workload (VGG16).}
    \label{fcomp}
\end{figure}

Fig.~\ref{fcomp} compares the hardware parameters and EDAP values for RRAM-based and SRAM-based architectures obtained through optimization for different objective functions. Overall, although SRAM-based designs tend to exhibit slightly lower energy consumption during processing, they suffer from higher latency due to the need for frequent weight swapping. The variation in on-chip area is also more significant in SRAM-based designs, as there are fewer constraints on how many weights must reside on-chip simultaneously. In general, RRAM-based architectures consistently achieve lower EDAP values compared to their SRAM-based counterparts.

For RRAM-based designs, where weight swapping is not used, the algorithm often converges to architectures with the maximum number of crossbar rows (512) and a relatively low number of columns, except in the case of area-based optimization ($f_j = A$), where both row and column sizes reach 512. Latency-focused optimization ($f_j = \max(L_w)$) tends to select designs with a smaller number of macros per tile but a higher number of tiles per router and per chip. In contrast, energy-focused optimization ($f_j = \max(E_w)$) converges toward architectures with many macros per tile and fewer tiles per chip.

In SRAM-based optimization, the algorithm typically converges to architectures with fewer crossbar rows and more columns compared to RRAM-based designs. As expected, area-focused optimization ($f_j = A$) produces the most compact architectures, but at the cost of increased weight swapping and thus higher latency. Energy-related optimization ($f_j = \max(E_w)$) prefers configurations with fewer tiles per router and fewer tile groups per chip. Meanwhile, latency-focused optimization ($f_j = \max(L_w)$) tends to select designs with larger on-chip area to minimize weight swapping, which in turn results in a higher number of tiles per chip.

\subsection{Ablation studies}
\label{ablationS}

\begin{figure}[t]
    \centering
    \includegraphics[width=\columnwidth]{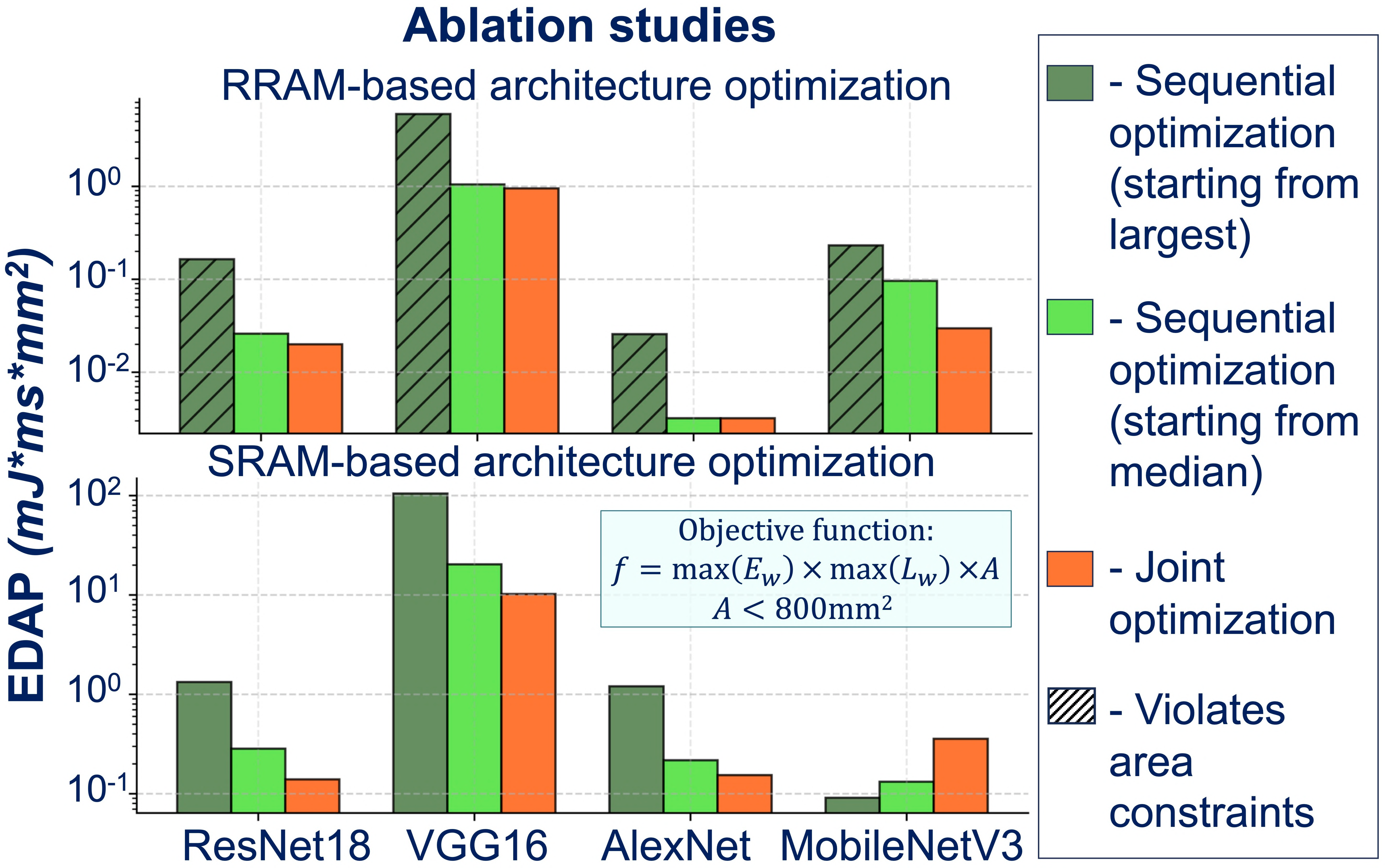}
    \caption{Ablation studies comparing the proposed joint hardware-workload optimization to a sequential hardware stack optimization.}
    \label{fablationF}
\end{figure}

We conducted ablation studies to compare the proposed joint hardware-workload co-optimization method with a sequential optimization approach for the hardware stack. Fig. \ref{fablationF} presents results for three distinct optimization strategies to minimize the energy-error-delay product, subject to a predefined area constraint: (1) joint optimization, (2) sequential optimization starting from the largest hardware configuration in the search space, and (3) sequential optimization starting from the median value of each parameter in the hardware design space. We explore two initialization cases for sequential optimization, as the choice of initial configuration significantly impacts the final performance. In each case, a single best-selected architecture is identified through hardware-workload co-optimization, and its performance is reported across different workloads. 
The sequential optimization is performed step by step across the hardware stack, starting from device-level parameters, then circuit-level, architecture-level, and finally system-level parameters. For RRAM-based designs, optimization begins at the device level and proceeds upward through the stack, while for SRAM-based designs, it starts at the circuit level due to the absence of device-level optimization. The optimization variables follow the categorization shown in Table \ref{t1}.

In both RRAM- and SRAM-based scenarios, joint hardware-workload optimization consistently outperforms sequential optimization. For RRAM, sequential optimization starting from the largest configuration fails to satisfy the area constraints, while starting from the median configuration satisfies them but results in inferior performance, especially for ResNet18, VGG16, and MobileNetV3. In SRAM, joint optimization delivers superior performance for all networks except MobileNetV3, with notably lower EDAP for the others. This effect occurs because, during the circuit-level optimization stage (while system and architecture parameters remain fixed), the algorithm selects crossbar sizes that are highly favorable for MobileNetV3 but suboptimal for the other networks. This choice reduces the objective value in the early stage but locks the search into a configuration that is difficult to correct later. When the optimization proceeds to the architecture level, these crossbar selections persist, and the sequential process cannot compensate for the earlier circuit-level bias. As a result, MobileNetV3 continues to achieve low EDAP, while the remaining networks exhibit higher scores. 
Overall, joint optimization achieves better total scores across all cases and remains independent of the initial hardware parameters, unlike sequential methods.

\subsection{Exploring the effects of RRAM non-idealities}
\label{nonideal}

Often, in IMC architectures based on non-volatile memories, design choices are influenced not only by performance metrics but also by device and hardware non-idealities that directly impact accuracy.
We demonstrate an application example of the proposed framework for hardware parameter optimization with non-ideal RRAM devices, where both hardware metrics and accuracy are incorporated into the objective function as $\max(E_w) \times \max(L_w) \times A/\prod_{i=1}^{4} \text{Acc}_{wi}$, where $\prod_{i=1}^{4} \text{Acc}_{wi}$ represents the product of accuracies across four workloads. 

To illustrate a practical scenario of a single hardware platform supporting multiple workloads and applications, we train 8-bit models using quantization-aware training on the following datasets: ResNet-18 on CIFAR-10, VGG16 on SVHN, AlexNet on Fashion-MNIST, and MobileNetV3 on CIFAR-100. The accuracies of the 8-bit baseline models are 94.88\%, 97.89\%, 93.5\%, and 70.03\%, respectively. The trained weights are then mapped to analog tiles using the AIHWKIT framework \cite{rasch2021flexible}, together with the RRAM model from \cite{wan2022compute} to capture noise and cycle-to-cycle variations. 
 {ReRAM conductance variability is modeled as additive Gaussian noise with a conductance-dependent standard deviation fitted to experimental data} \cite{wan2022compute}.  {For a target conductance $g_t$,}
\begin{equation}
g = g_t + \sigma(g_t)\,\varepsilon,
\qquad
\varepsilon \sim \mathcal{N}(0,1),
\end{equation}
 {where $\sigma(g_t)$ is a conductance-dependent standard deviation modeled using a fourth-order polynomial fit to experimental data from} \cite{wan2022compute}  {as a function of the normalized conductance} \cite{rasch2021flexible}.
We further include the effects of IR-drop, 8-bit quantization of the converters, and 1\% output noise following \cite{rasch2021flexible}.  {IR-drop is modeled as an approximate resistive interconnect effect at the crossbar level, 8-bit DAC/ADC quantization is captured via uniform bounded discretization, and output noise is modeled as additive Gaussian output-referred noise.}
It should be noted that the models were not retrained using a hardware-aware training approach \cite{rasch2023hardware}, since such retraining is time-consuming, would considerably increase the search time, and falls outside the scope of this work. Consequently, we observe an accuracy drop relative to the 8-bit baselines due to device- and circuit-level non-idealities. As shown in \cite{rasch2023hardware}, hardware-aware retraining can substantially recover accuracy.

\begin{figure}[t]
    \centering
    \includegraphics[width=\columnwidth]{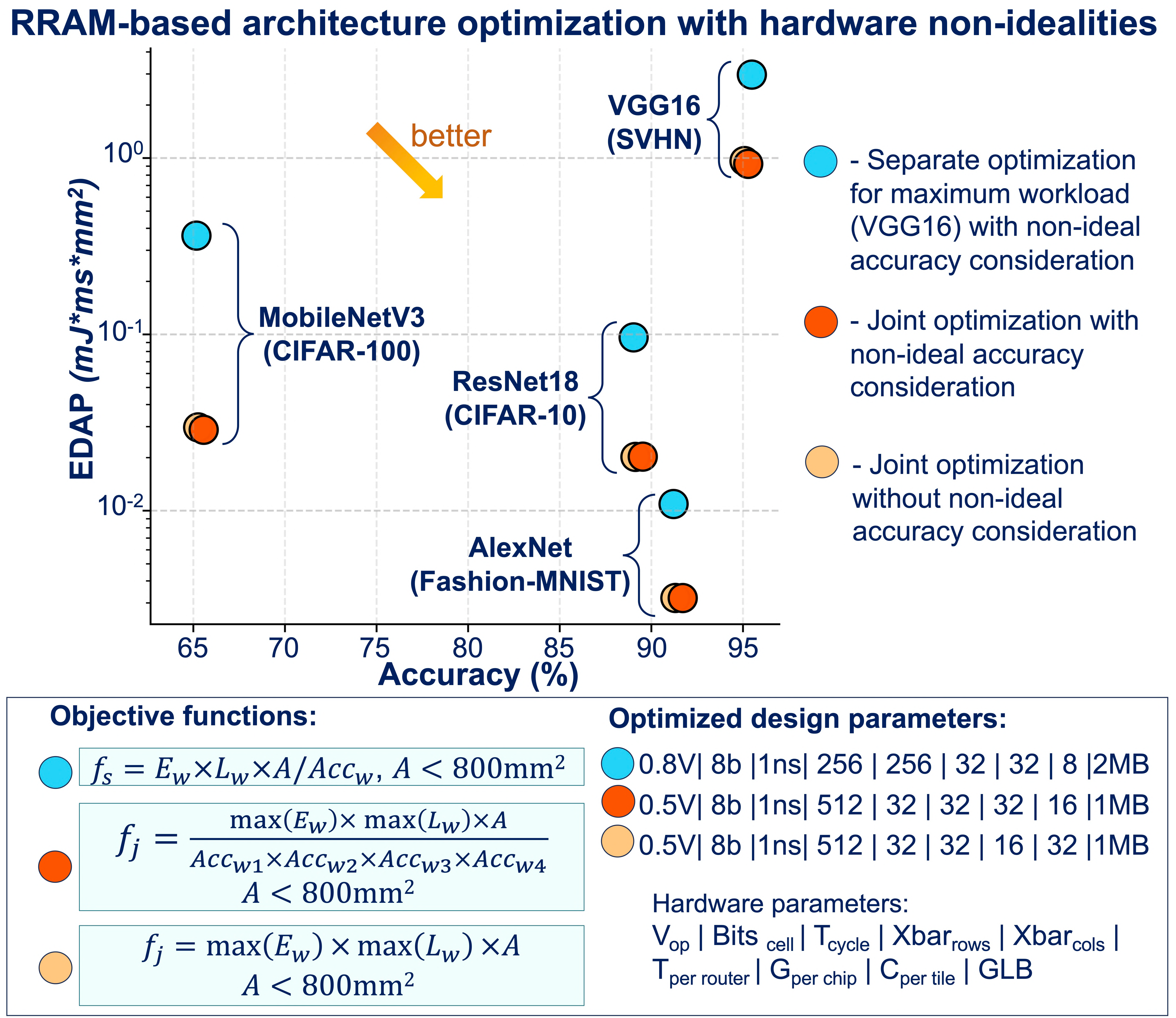}
    \caption{Experiments considering RRAM hardware non-idealities.}
    \label{nonidealitiesF}
\end{figure}

The results of this experiment are presented in Fig. \ref{nonidealitiesF}, comparing EDAP and accuracy for designs obtained via joint optimization and optimization for the largest workload considering non-ideal accuracy in the objective function, and joint optimization targeting EDAP only.
Accuracy under non-idealities is reported as the average over 30 iterations with random cycle-to-cycle variations. Similar to previous experiments, joint optimization consistently achieves better designs compared to maximum-workload-only optimization. Moreover, both joint optimization experiments converge to nearly the same architecture, regardless of whether hardware non-idealities are explicitly considered. This is because short-term noise (cycle-to-cycle variation) has a more significant impact on accuracy than IR-drop, which primarily depends on crossbar sizes.

\subsection{Exploring trade-offs between hardware efficiency and fabrication cost}
\label{Stradeoffs}

While existing works on hardware design space exploration and parameter optimization focus on a single CMOS technology node and aim to optimize hardware parameters within a framework with a fixed CMOS technology \cite{negi2022nax, yang2021multi, moitra2023xpert}, they often overlook the practical trade-offs between fabrication cost and hardware performance. In this work, we address this limitation by incorporating CMOS technology node selection as an additional optimization parameter. Our goal is to identify not only the optimal hardware configurations for various neural network workloads, but also the most cost-effective CMOS technology node for fabricating the corresponding chip. This joint optimization allows us to balance hardware performance and fabrication cost, enabling more practical and efficient hardware design for real-world deployment.

To enable cost-aware hardware design, we integrate fabrication cost into the objective function, formulating the optimization problem as $\max(E_w)\times\max(L_w)\times Cost$, $\mathrm{s.t.} \, A \leq A_{constr}$, 
where $A_{constr}=800$mm$^2$ and $Cost = \alpha \times A$, with $\alpha$ representing the normalized fabrication cost per mm$^2$ for the selected technology node. Since fabrication cost is directly proportional to the chip area, we do not include the area term separately in the objective function. 

\begin{table}[t!]
\caption{Cost calculation parameters and voltage range for different technologies.}
\resizebox{\columnwidth}{!}{%
\begin{tabular}{|l|c|c|c|c|}
\hline

\multicolumn{1}{|c|}{\textbf{Technology}} &
\begin{tabular}[c]{@{}c@{}}
  \textbf{Average} \\ \textbf{wafer cost}\\ \textbf{(USD)}
\end{tabular} &
\textbf{Yield} &
\begin{tabular}[c]{@{}c@{}}
  \textbf{Average cost $\alpha$} \\ \textbf{per mm$^2$} \\ \textbf{(normalized to 32nm)}
\end{tabular} &
\begin{tabular}[c]{@{}c@{}}
  \textbf{Voltage range}\\ \textbf{for} \\ \textbf{simulation}
\end{tabular} \\ \hline

\textbf{90nm}                             & 1651.5                                                                         & 90-95\%        & 0.413                                                                                            & 0.95-1.3V                                                                          \\ \hline
\textbf{65nm}                             & 1939.0                                                                          & 90-95\%        & 0.477                                                                                            & 0.85-1.2V                                                                          \\ \hline
\textbf{45nm}                             & 2237.5                                                                         & 80-90\%        & 0.606                                                                                            & 0.75-1.1V                                                                          \\ \hline
\textbf{32nm}                             & 3500.0                                                                           & 70-90\%        & 1                                                                                                & 0.65-1.0V                                                                            \\ \hline
\textbf{22nm}                             & 4338.5                                                                         & 70-90\%        & 1.282                                                                                            & 0.65-1.0V                                                                            \\ \hline
\textbf{14nm}                             & 4492.0                                                                           & 60-80\%        & 1.498                                                                                           & 0.55-0.9V                                                                          \\ \hline
\textbf{10nm}                             & 5600.0                                                                           & 50-70\%        & 2.243                                                                                            & 0.5-0.85V                                                                          \\ \hline
\textbf{7nm}                              & 9291.5                                                                         & 50-70\%        & 3.871                                                                                            & 0.45-0.8V                                                                          \\ \hline
\end{tabular}
}
\label{t3}
\end{table}

The selected technology nodes are shown in Table \ref{t3}. Technology nodes above 45nm typically feature lower wafer costs, fewer processing steps, and higher yields, making them mature and cost-effective options. Nodes such as 28nm and 32nm represent well-established processes that offer a good trade-off between performance and cost, and are widely used as reference points. Accordingly, we normalize the cost in our simulations to the 32nm CMOS technology node. In contrast, the 22nm node involves more complex processing steps, leading to increased fabrication costs. More advanced nodes, such as 7nm to 14nm, utilize FinFET transistors, which require a significantly more intricate manufacturing process. As the technology node shrinks, fabrication costs rise due to higher mask complexity and lower manufacturing yields.

\begin{figure}[t!]
    \includegraphics[width=\columnwidth]{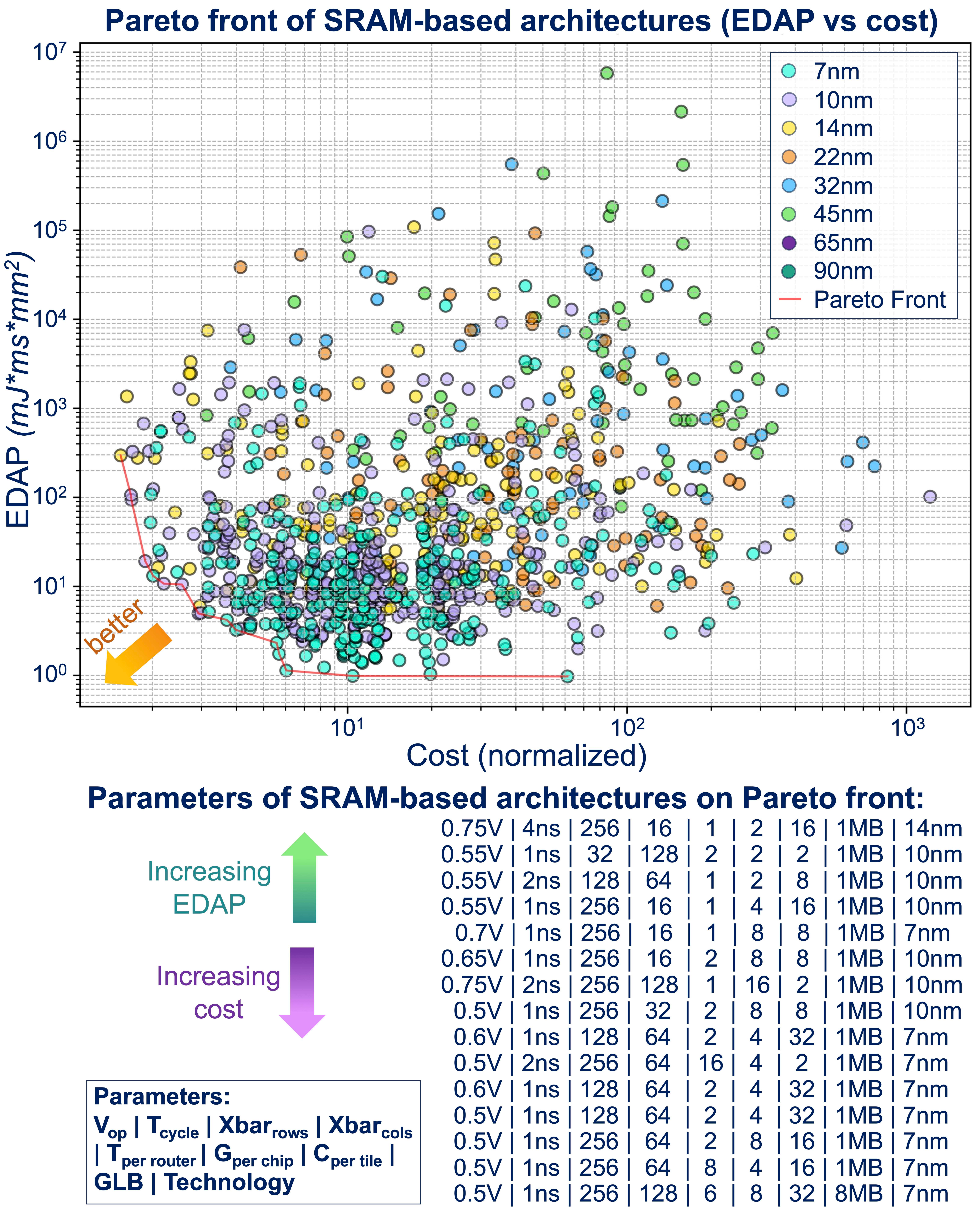}
    \caption{Trade-offs between EDAP and fabrication cost and corresponding Pareto front for SRAM-based hardware-workload-technology design optimization.}
    \label{f4}
\end{figure}

\begin{figure*}[t!]
    \includegraphics[width=\textwidth]{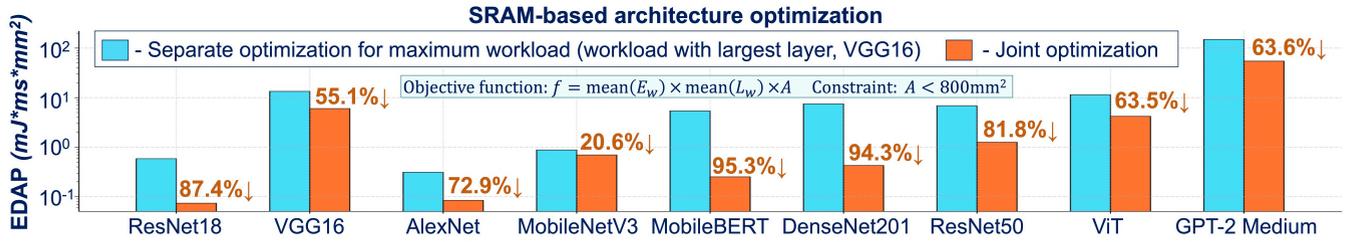}
    \caption{Scalability and generalization of the proposed joint hardware-workload co-optimization method across multiple workloads and network types, with a comparison to optimization targeting the maximum workload (i.e., the workload with the largest layer, in the context of weight-swapping SRAM-based IMC).}
    \label{fnewnetweoks}
\end{figure*}

To calculate the cost per mm$^2$ for different technology nodes, we relied on publicly available data for the cost of 300mm wafers from \cite{khan2020ai, anysilicon_wafer_cost}. We consider substantial yield variations, especially in large dies such as accelerator chips, where yield drops significantly \cite{anandtech_tsmc_5nm}. The average wafer cost $C_{avg}$ used in our analysis is summarized in Table~\ref{t3}, averaged from the sources mentioned above.
Assuming a 300mm wafer diameter, the total wafer area is approximately $70685$ mm$^2$, of which around 95\% is considered usable \cite{johnson_cmos_cost}, resulting in an effective area $A_e \approx 70000$ mm$^2$. The functional wafer area is then calculated as $A_{f} = A_e \times \mathrm{yield}$. We use average yield values, as precise, open-source yield data is not consistently available and varies across technology nodes and fabrication foundries.
The average cost per mm$^2$ is calculated as $C_{per\, mm^2}=\frac{C_{avg}}{A_{f}}$, and is subsequently normalized to the 32nm CMOS technology node, represented as the cost scaling factor $\alpha$ in Table~\ref{t3}.
Although exact values for the cost per mm$^2$ are scarce in the literature, our estimated relative costs follow an exponential trend with decreasing technology node size, which aligns with trends reported in \cite{johnson_cmos_cost}, thereby supporting the validity of our normalization approach.

In addition, we adjust the operating voltage range according to the selected CMOS technology node; therefore, the $\mathrm{V_{op}}$ parameter varies across technologies. The corresponding ranges are summarized in Table~\ref{t3}.

We perform the hardware-workload-technology co-optimization using an SRAM-based design, chosen for its maturity and more reliable cost estimation compared to RRAM. The optimization results, jointly considering hardware performance and fabrication cost, are shown in Fig.~\ref{f4}, which presents all feasible architectures satisfying the area constraints. Designs based on 65 nm and 90 nm CMOS are not shown on the graph, as they do not satisfy the area constraints.
From these results, the Pareto front is constructed to illustrate the EDAP-cost trade-off, and the parameters of the optimized architectures on the Pareto front are also indicated in Fig.\ref{f4}.
These optimal designs are primarily based on 7nm to 14nm CMOS technologies, where 7nm designs are typically located on the lower EDAP but higher cost end of the Pareto front, while 10-14nm designs offer reduced cost but slightly higher EDAP. 
The lower-cost designs on the Pareto front (primarily 10-14nm nodes) generally achieve smaller area footprints by reducing the number of tiles per router and the number of tile groups per chip, rather than minimizing the number of macros per tile. The most optimal trade-offs between performance and fabrication cost are predominantly achieved by 10nm designs, and these architectures are not the smallest in terms of component count (e.g., tile groups per chip). On the higher-cost end of the Pareto front with the smallest EDAP metrics (mostly based on 7nm technology), the designs also do not necessarily converge to architectures with the minimal number of components. Instead, they contain a larger number of processing elements, which leads to fewer weight swaps and, consequently, lower energy consumption and latency during data transmission between the IMC chip and DRAM.

\subsection{Scalability and generalization of the proposed approach}
\label{Sgeneralization}

To assess the scalability and generalization of the proposed joint hardware-workload co-optimization, we performed an additional experiment on SRAM-based architectures with an expanded search space of diverse workloads: ResNet18, VGG16, AlexNet, MobileNetV3, MobileBERT, DenseNet201, ResNet50, Vision Transformer (ViT), and GPT-2 Medium. The experiment targeted acceleration on SRAM-based in-memory computing hardware using weight-swapping at 32 nm CMOS, consistent with prior experiments and modeled as in \cite{andrulis2024cimloop} for transformer-based workloads. Since GPT-2 Medium dominates energy and latency, the original objective function in Eq.~\ref{eq:o1} would bias optimization toward this model. To avoid this, we redefined the objective to use mean energy and latency across workloads, rather than maxima  {(as shown in Section} \ref{subsec_newAggregation}), under the same area constraints.
It should be noted that for transformer-based models, we focus on optimizing medium-sized workloads, as pure SRAM-based IMC hardware is less suitable for very large models. In such cases, alternative platforms, such as hybrid designs combining IMC with digital computing systems \cite{gao2024imi} or IMC with computation-centric accelerators \cite{he2025papi}, offer more practical solutions.

Fig.\ref{fnewnetweoks} presents the results of this experiment, where the proposed joint optimization approach is compared against optimization targeting the largest workload. Since the experiments are conducted on SRAM-based hardware with weight-swapping, the largest workload is defined by the layer containing the highest number of weights that must fit into the IMC hardware, rather than by overall model size. Under this definition, VGG16 is considered the largest workload, with a single layer containing $8.2\times 10^8$ parameters, compared to $4.1\times 10^8$ in the largest layer of GPT-2 Medium, even though GPT-2 Medium is the larger model overall. The simulation results in Fig.\ref{fnewnetweoks} show the same trend as earlier experiments: the proposed joint optimization consistently outperforms optimization targeting only the largest workload.  {In addition, the total search time is approximately 99 hours on 64 CPU cores, of which around 35 hours are attributed to the sampling phase. This is consistent with the results reported in Section} \ref{subruntime},  {where the sampling overhead is observed to be approximately 30\%.}  These results highlight the scalability and adaptability of the framework, demonstrating that joint hardware-workload co-optimization can be effectively extended to diverse workloads with heterogeneous computational and memory characteristics, including transformer-based models.

\vspace{1cm}

\section{Discussion}
\label{Sdisc}

\subsection{Why is joint workload-hardware co-optimization important?}

In most practical applications, multiple workloads with varying structures and sizes must be executed on a single hardware platform. Consequently, an IMC accelerator should efficiently support more than one neural network model. As shown in \cite{krestinskaya2024towards}, weight-stationary designs optimized for smaller models often fail to accommodate more complex networks, whereas designs tuned for the largest workload lead to inefficient configurations for smaller or structurally different models. This demonstrates the limitations of single-workload-driven design and motivates the need for a more holistic approach.  {The results in Section}~\ref{SRA}  {and Section}~\ref{Sgeneralization}  {demonstrate that joint optimization, compared to optimization targeting only the largest workload, reduces EDAP by up to 76.2\% and 95.5\% when optimizing across four-workload and nine-workload sets, respectively} (Fig.\ref{f1p2} and Fig.\ref{fnewnetweoks}).  {Even though this improvement varies across different models, the proposed approach consistently outperforms the baseline method.} 
Our joint optimization strategy addresses this challenge by balancing performance across diverse workloads, thereby enabling more efficient and generalizable hardware architectures. Furthermore, Section~\ref{ablationS}  {and Fig.} \ref{fablationF} demonstrate that jointly optimizing the full stack of hardware parameters across multiple levels of the hierarchy  {in a single optimization run} achieves superior results compared to sequential step-by-step optimization  over different levels of the hardware hierarchy. Sequential optimizations are highly dependent on the initial set of fixed parameters that are not optimized in the current step and tend to get stuck in local minima, producing designs with worse metrics compared to the joint approach. In some cases, particularly for RRAM-based optimization, sequential optimization can even violate area constraints. Therefore, joint optimization is a more robust and effective approach for achieving efficient designs.

\subsection{Can performance loss of generalized designs be reduced?}

It is well understood that moving toward generalized architectures capable of supporting various workloads often leads to performance degradation, especially when compared to hardware optimized for a single, specific neural network model. This raises a critical question: can the performance gap between generalized and workload-specific hardware be effectively minimized? In Section~\ref{SRC} and Fig.~\ref{f3}, we demonstrate that the proposed four-phase genetic algorithm-based joint optimization significantly reduces this gap, and consistently produces generalized designs with EDAP scores (or other objective-specific metrics) closer to those of workload-specific optimization, outperforming other baseline methods.  {Although the generalized architecture cannot achieve the same performance as an architecture optimized for a specific workload, and in some cases the algorithm trades off performance among different workloads} (Fig.~\ref{f3} (f-g)),  {the overall performance is improved significantly.}
 {In addition, Section}~\ref{SRB}  {and Fig.}~\ref{f2}  {demonstrate that the proposed algorithm significantly contributes to the overall improvement in the EDAP scores of the optimized designs. It improves convergence stability and optimization quality by incorporating a Hamming-distance-based sampling strategy, which enhances diversity in the initial population, mitigates premature convergence, and leads to more consistent outcomes across independent runs} (Section~\ref{SRB}).
Overall, we demonstrated that the performance loss associated with generalized hardware can be effectively minimized.

\subsection{How to benefit from the framework in practical scenarios?}

We demonstrate the framework’s applicability in several practical scenarios: optimizing RRAM-based IMC hardware under device non-idealities, enabling hardware-workload-technology co-optimization for SRAM-based systems, and scaling to diverse workloads.
 As demonstrated in Section~\ref{SRD} and Fig.~\ref{fcomp}, the framework can be used to compare the performance of different memory technologies, and it can be adapted to evaluate the hardware performance alongside performance accuracy under non-ideal device and circuit conditions (Section~\ref{nonideal} and Fig. \ref{nonidealitiesF}).
 As shown in Section~\ref{Stradeoffs} and Fig.~\ref{f4}, it supports joint hardware-workload-technology co-optimization across multiple CMOS nodes, enabling exploration of configurations that balance performance and fabrication cost while quantifying trade-offs between efficiency and cost-effectiveness. Fig.~\ref{f4} demonstrates an example of such an application for SRAM-based optimization, including the cost calculation method, as well as the Pareto front and corresponding design choices for EDAP-cost trade-offs.
The framework can be extended to support diverse memory technologies, system models, and evaluation constraints, and can be customized with designer-specific parameters or cost models. As shown in Section~\ref{Sgeneralization}, it scales to larger and more diverse workloads, including medium-sized transformer models. This adaptability makes the framework well-suited for guiding hardware design choices across a wide range of realistic IMC scenarios and for diverse design exploration and deployment cases.

\subsection{How to improve the runtime of the algorithm?}
\label{DiscussionRuntime}

In addition, an important aspect to consider is the search runtime.  {It should be noted that the introduction of enhanced sampling and the modified four-phase algorithm increases the runtime by approximately 30\%. Depending on the population size, simulations using the proposed algorithm require up to 19 h on 64 CPU cores for the 4-workload experiment, and up to 99 h for the 9-workload experiment involving large workloads.}  Despite the increased search time, the significant improvement in hardware performance justifies the additional computational cost, especially when targeting generalized architectures.

The total runtime can be reduced or increased by varying the population parameters $P_H$ and $P_E$. In this work, we select relatively large population sizes to ensure that the initial population consists of well-performing architectures. The number of generations $G$ can be adjusted to further optimize runtime. For example, reducing the number of generations in the transition and convergence phases can significantly decrease the total search time with minimal impact on the performance of the final hardware design.  {In addition, another possible approach to reduce the number of generations is to monitor the convergence of the algorithm during the search and apply early stopping when convergence is reached, rather than running through all generations in each phase of the four-phase search.}

 {To significantly reduce the overall search time, hardware metric prediction models could be incorporated by training dedicated predictors in place of explicit hardware estimation for each sampled design. However, such an approach must be carefully designed to ensure high prediction accuracy. Unlike predictors used for neural network accuracy estimation, which often capture only general performance trends rather than exact accuracy} \cite{krestinskaya2025cimnas},  {hardware metric prediction requires substantially higher accuracy, as even small errors can significantly affect optimization outcomes. }

\subsection{How can the framework be extended and what are the next steps forward?}

\label{futureDisc}

\subsubsection{ {Optimization of neural network model parameters}}

The proposed hardware-workload co-optimization scenario can be further extended to include the optimization of neural network model parameters for each supported workload. This would address the research gap highlighted in Fig.~\ref{fmotive} for three-level optimization. Such an extension may involve accuracy-driven optimization of model components, such as neural network blocks and layer sizes \cite{krestinskaya2024neural}, to further enhance hardware efficiency without compromising performance accuracy. However, achieving this joint optimization poses additional challenges due to the significantly larger and more complex search space, requiring further advancements in the underlying optimization methodology.
Compared to hardware-only optimization to support multiple workloads, the search space size for joint model-hardware co-optimization can increase dramatically, from approximately $10^7$ to $10^{84}$ for a single workload \cite{sun2023gibbon}. Extending this co-optimization to support multiple workloads, by simultaneously tuning model parameters and generalized IMC hardware configurations, introduces an additional layer of complexity. This requires the development of new co-optimization strategies capable of efficiently managing such expansive design spaces.

\subsubsection{ {Process/voltage/temperature- and reliability-aware optimization}}

\label{Subsec_PTV}

 {As a future direction, the proposed framework can be combined with complementary modeling and optimization tools to enable end-to-end system-level optimization that accounts for process, voltage, and temperature (PVT) effects. Thermal- and reliability-aware objectives or constraints can be incorporated into the architecture-level exploration as the required technology-specific parameters become available. Existing thermal-aware mapping and optimization techniques} \cite{smagulova2024thermal}  {can then be integrated to jointly optimize accuracy, energy, latency, area, thermal behavior, and reliability in an end-to-end manner.
However, this integration introduces new challenges related to efficiently incorporating thermal and reliability estimation methods without significantly increasing the overall search time.}

\subsubsection{ {Large language and reasoning models support}}

 {Extending the framework to support very large neural network models, including transformer-scale architectures with billions of parameters, requires the use of hierarchical model abstractions and block-level representations to ensure scalable exploration without explicitly instantiating all parameters. Hardware-workload co-optimization for such models must account for layer-wise heterogeneity, as different layers exhibit distinct compute and memory demands, as well as overall on-chip memory optimization and inter-layer communication overheads.
However, supporting models at this scale substantially increases both the design-space complexity and the cost of performance evaluation, motivating the use of surrogate modeling, progressive refinement strategies, and multi-fidelity optimization techniques to maintain computationally feasible search complexity.}

\subsubsection{ {Multi-chip and multi-die systems support}}

 {Another possible extension of the proposed framework is support for multi-chip and multi-die systems, where large workloads are distributed across multiple accelerator instances or chiplets} \cite{shao2019simba, srivastava2025corsair}.  {This requires incorporating interconnect bandwidth, communication latency, synchronization overheads, and data placement strategies into the optimization loop. Joint optimization of hardware parameters, workload mapping, and partitioning strategies would enable efficient scaling beyond single-chip constraints.
However, incorporating multi-chip considerations introduces additional system-level design choices, such as topology selection and communication scheduling, which significantly increase the dimensionality of the search space and require new optimization strategies.}

\subsubsection{ {Training workloads support}}

 {Another important direction is extending the framework to explicitly support hardware for training workloads, rather than focusing primarily on inference. Training introduces additional challenges, including weight update dynamics, higher precision requirements, increased on-chip or off-chip storage demands for batch training, and substantially different compute and memory access patterns. Incorporating training-aware objectives requires modeling forward and backward passes, gradient accumulation, and weight update operations. In addition, efficient approximations are needed, such as evaluating hardware using reduced workloads (fewer batches or smaller models) as proxies for training performance.
Training-aware co-optimization can substantially increase evaluation cost and modeling complexity, motivating the development of workload-aware abstractions and efficient approximation techniques that capture dominant training bottlenecks while keeping the optimization computationally feasible.}

\section{Conclusion}
\label{Sconclu}

 {This paper presented a framework for joint hardware-workload co-optimization of generalized IMC architectures that support multiple neural network workloads while considering a wide range of hardware parameters. The four-phase genetic algorithm with Hamming-distance-based sampling was proposed to improve convergence behavior and optimization quality, leading to more consistent and efficient hardware designs. 
The results demonstrated that the proposed approach effectively reduced the performance gap between generalized architectures and single-workload-optimized designs, while supporting diverse memory technologies, workloads, and design scenarios. Overall, the framework proved to be a robust and practical tool for efficient exploration and design of multi-workload IMC accelerators.}

 {Future work can extend the framework toward full model-hardware co-optimization for multiple workloads and introduce methods to efficiently manage the substantially larger associated search spaces. Additional directions include incorporating process, voltage, temperature, and reliability awareness into the optimization loop, enabling scalable support for large language and reasoning models through hierarchical abstractions and surrogate modeling, extending the framework to multi-chip and multi-die systems with explicit modeling of communication and synchronization overheads, and supporting training workloads through training-aware objectives and efficient performance approximations.}

\bibliographystyle{ieeetr}
\bibliography{references.bib}

\begin{IEEEbiography}[{\includegraphics[width=1in,height=1.25in,clip,keepaspectratio]{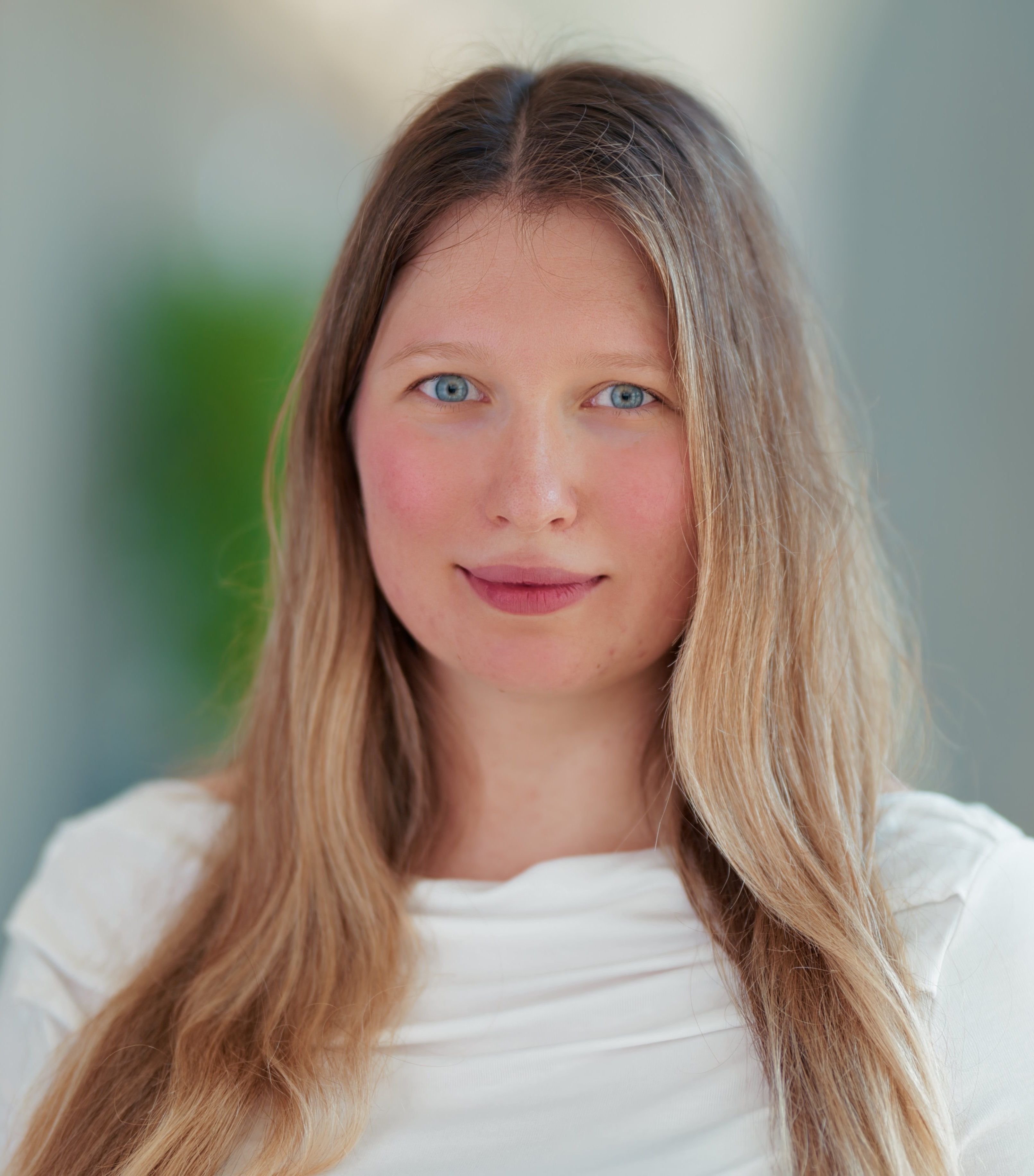}}]{Olga Krestinskaya}
received her Ph.D. degree from King Abdullah University of Science and Technology (KAUST), Saudi Arabia, in 2025, where she is currently a Postdoctoral Fellow. Her research focuses on software-hardware co-design for in-memory computing (IMC) architectures and AI hardware. She has published over 50 peer-reviewed works, including high-impact journal articles, conference papers, and book chapters, covering analog memristive neural networks, mixed-signal circuit-level implementations of IMC architectures, quantized neural networks, and brain-inspired algorithms, with a focus on developing energy-efficient and scalable IMC hardware for AI applications.
Dr. Krestinskaya is the recipient of the 2019 IEEE CASS Predoctoral Award, Erasmus Student Mobility Scholarship, KAUST Dean's Scholarship, the 2025 Web of Talents STEM Award (1st place), and multiple KAUST Dean’s Awards. Her work was recognized with the Best Poster Award at the 2nd Nature Conference on Neuromorphic Computing (2024), and she was shortlisted for the prestigious Rising Stars Women in Engineering Workshop (Asian Deans’ Forum 2024). She also serves as a peer reviewer for many journals and conferences.
\end{IEEEbiography}

\begin{IEEEbiography}[{\includegraphics[width=1in,height=1.25in,clip,keepaspectratio]{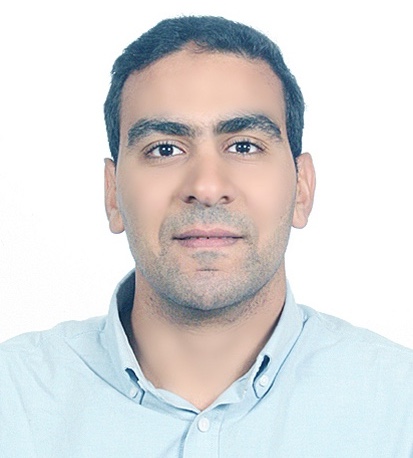}}]{Mohammed E. Fouda} received a B.Sc. degree (Hons.) in Electronics and Communications Engineering and an M.Sc. degree in Engineering Mathematics from the Faculty of Engineering, Cairo University, Cairo, Egypt, in 2011 and 2014, respectively. Fouda received a Ph.D. degree from the University of California, Irvine, USA, in 2020. Dr. Fouda worked as an applied research lead at Rain Neuromorphics Inc till June 2025. Currently, he works as the chief technology officer at Compumacy for Artificial Intelligence Solutions.  Dr. Fouda has published more than 200 peer-reviewed journal and conference papers, one Springer book, and three book chapters. His H-index is 35, and he has been cited more than 4300 times. His research interests include analog AI hardware and algorithms, neuromorphic circuits and systems, brain-inspired computing, and analog circuits. He serves as a peer reviewer for many prestigious journals and conferences. He also serves as an associate editor in many journals in addition to serving as a technical program committee member in many conferences. He was the recipient of the best paper award in ICM for 2013 and 2020 and the Broadcom Foundation fellowship for 2016/2017.
\end{IEEEbiography}

\begin{IEEEbiography}[{\includegraphics[width=1in,height=1.25in,clip,keepaspectratio]{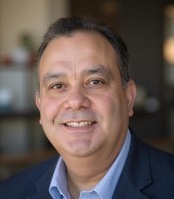}}]{Ahmed M. Eltawil} (Senior Member, IEEE) received the bachelor’s and master’s degrees from Cairo University, in 1997 and 1999, respectively, and the doctorate degree from the University of California, Los Angeles, in 2003. He is a Professor and an Associate Dean for Research at the Computer, Electrical, and Mathematical Sciences and Engineering (CEMSE) Division, King Abdullah University of Science and Technology (KAUST). Previously, he was a Professor of Electrical Engineering and Computer Science at the University of California, Irvine (UCI) from 2005 to 2021. At KAUST, he established the Communication and Computing Systems Laboratory (CCSL) to conduct research on efficient architectures for computing and communications systems, with a particular focus on mobile wireless systems. His research interests encompass various application domains, such as low-power mobile systems, machine learning platforms, sensor networks, body area networks, and critical infrastructure networks. He served as a distinguished lecturer for IEEE COMSOC during the 2023/24 term. Additionally, he holds senior membership with the National Academy of Inventors, USA. He received several recognitions and awards, including the US National Science Foundation CAREER award, the 2021 “Innovator of the Year” award by the Henry Samueli School of Engineering at the University of California, Irvine, and two United States Congress certificates of merit, among other recognitions. He has served in numerous editorial roles over the years, as well as an expert reviewer for national and international funding agencies and review boards.
\end{IEEEbiography}

\begin{IEEEbiography}[{\includegraphics[width=1in,height=1.25in,clip,keepaspectratio]{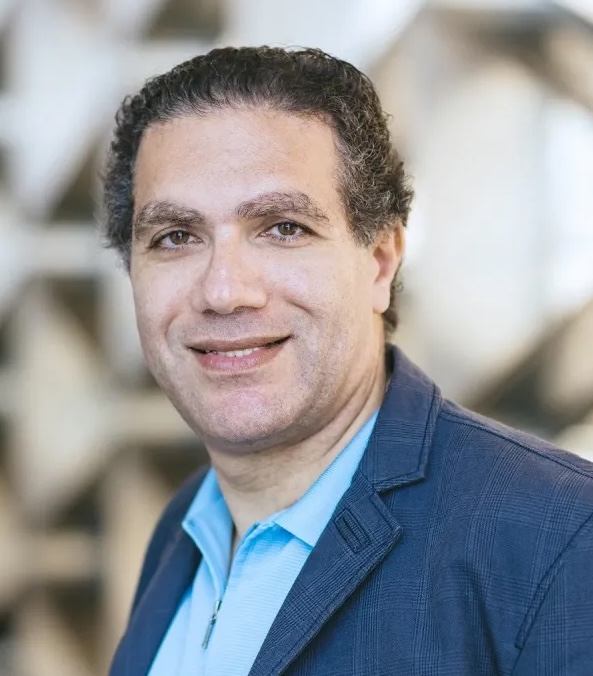}}]{Khaled N. Salama}  (Senior Member, IEEE) received the B.S. (Hons.) degree from the Department of Electronics and Communications, Cairo University, Giza, Egypt, in 1997, and the M.S. and Ph.D. degrees from the Department of Electrical Engineering, Stanford University, Stanford, CA, USA, in 2000 and 2005, respectively. He was an Assistant Professor with the Rensselaer Polytechnic Institute, Troy, NY, USA, from 2005 to 2009. In 2009, he joined the King Abdullah University of Science and Technology, Thuwal, Saudi Arabia, where he was the Founding Program Chair until 2011 and Associate Dean between 2019 and 2022, where he is currently a Professor. He has authored 450 articles and holds 50 patents on low-power mixed-signal circuits for intelligent fully integrated sensors and nonlinear electronics, in particular memristor devices. His research on CMOS sensors for molecular detection has been funded by the National Institutes of Health and the Defense Advanced Research Projects Agency.,Dr. Salama was a recipient of the Stanford-Berkeley Innovators Challenge Award in Biological Science.
\end{IEEEbiography}

\EOD

\end{document}